\PassOptionsToPackage{dvipsnames}{xcolor}
\documentclass[desactivate]{aa}
\graphicspath{{figures/}}

\usepackage[colorlinks=true, linkcolor=blue, citecolor=blue, filecolor=blue, urlcolor=blue]{hyperref}
\usepackage{lipsum}
\usepackage{physics}
\usepackage{multirow}
\usepackage{xspace}
\usepackage{natbib}
\usepackage{fontawesome5}
\usepackage{wrapfig}
\usepackage[figuresright]{rotating}
\usepackage{fontawesome5}

\usepackage[hang,flushmargin]{footmisc}

\begin{document}

\title{The Asteroseismic Imprints of Mass Transfer}
\subtitle{A Case Study of a Binary Mass Gainer in the SPB Instability Strip}
% \titlerunning{Imprints of mass transfer on asteroseismic signals}

\author{Tom Wagg\thanks{tomjwagg@gmail.com} \fnmsep \inst{1,2,3} \orcid{0000-0001-6147-5761}
        \and Cole Johnston \inst{1,4,5} \orcid{0000-0002-3054-4135}
        \and Earl P. Bellinger \inst{1,6,7} \orcid{0000-0003-4456-4863}
        \and Mathieu Renzo \inst{8,3} \orcid{0000-0002-6718-9472}
        \and\\Richard Townsend \inst{9} \orcid{0000-0002-2522-8605}
        \and Selma E. de Mink \inst{1} \orcid{0000-0001-9336-2825}}

\institute{Max-Planck-Institut für Astrophysik, Karl-Schwarzschild-Straße 1, 85741 Garching, Germany
           \and Department of Astronomy, University of Washington, Seattle, WA, 98195
           \and Center for Computational Astrophysics, Flatiron Institute, 162 Fifth Ave, New York, NY, 10010, USA
           \and Department of Astrophysics, IMAPP, Radboud University Nijmegen, PO Box 9010, 6500 GL Nijmegen, Netherlands
           \and Institute of Astronomy, KU Leuven, Celestijnenlaan 200D, 3001 Leuven, Belgium
           \and Department of Astronomy, Yale University, CT, 06511, USA
           \and Stellar Astrophysics Centre, Aarhus University, Aarhus, Denmark
           \and Steward Observatory, University of Arizona, 933 N. Cherry Avenue, Tucson, AZ 85721, USA
           \and Department of Astronomy, University of Wisconsin-Madison, 475 N Charter St, Madison, WI 53706, USA}

\date{Received \today{}}

\abstract{
    We present new simulations investigating the impact of mass transfer on the asteroseismic signals of slowly pulsating B stars. We use \texttt{MESA} to simulate the evolution of a binary star system and \texttt{GYRE} to compute the asteroseismic properties of the accretor star. We show that, compared to a single star of the same final mass, a star that has undergone accretion (of non-enriched material) has a significantly different internal structure, evident in both the hydrogen abundance profile and Brunt–Väisälä frequency profile. These differences result in significant changes in the observed period spacing patterns, implying that one may use this as a diagnostic to test whether a star's core has been rejuvenated as a result of accretion.
    We show that it is essential to consider the full multimodal posterior distributions when fitting stellar properties of mass-gainers to avoid drawing misleading conclusions. Even with these considerations, stellar ages will be significantly underestimated when assuming single star evolution for a mass-gainer. We find that future detectors with improved uncertainties would rule out single star models with the correct mass and central hydrogen fraction.
    Our proof of principle analysis demonstrates the need to further investigate the impact of binary interactions on stellar asteroseismic signals for a wide range of parameters, such as initial mass, amount of mass transferred and the age of the accretor star at the onset of mass transfer.
}


\maketitle

% {\hypersetup{linkcolor=black}\listoftodos}

\section{Introduction} \label{sec:intro}

The majority of stars are born in binaries and multiple star systems \citep[e.g.][]{Duchene+2013:2013ARA&A..51..269D,Moe+2017,Offner+2023:2023ASPC..534..275O}, a large subset of which will exchange mass at some point in their lifetime \citep[e.g][]{Podsiadlowski+1992:1992ApJ...391..246P,Sana+2012,deMink+2014}. However, mass transfer, both the process itself and the impact it has on the component stars, is still highly uncertain. Specifically, there are large uncertainties in how much mass and angular momentum is transferred and what part is lost from the system \citep{Packet+1981,deMink+2007:2007A&A...467.1181D,Renzo+2021} and how the accretor star adjusts to the incoming mass \citep{Hellings1983,Braun+1995,Cantiello+2007,Staritsin+2019,Renzo+2023,Lau+2024:2024arXiv240109570L}. These uncertainties in understanding the process of mass transfer result in uncertainties in evolutionary calculations and predictions, such as the rate of formation of close double compact objects \citep{Toonen+2013,Marchant2021+, vanSon+2022:2022ApJ...940..184V}, stripped stars \citep{Crowther2007,Heber2016,Gotberg+2020}, X-ray binaries \citep{Fragos+2013:2013ApJ...764...41F}, and gravitational wave sources \citep[e.g.][]{Dominik+2015:2015ApJ...806..263D,Belczynski+2002:2002ApJ...572..407B,Broekgaarden+2022,Iorio+2023}.

Asteroseismology probes the internal structure of stars through the analysis of stellar pulsations \citep{Aerts+2010} and so may hold the key to directly probing how accretor stars adjust to gaining mass \citep[e.g.][]{Renzo+2021}. %It is well established that asteroseismology can be used to estimate precise stellar masses, radii and ages based on stellar oscillation modes \citep{Aerts+2010}.
In particular, high-order gravity ($g$) mode pulsations carry information about the deep radiative interiors of stars and the boundary between the convective core and radiative envelope. Main-sequence stars that exhibit $g$-modes include F-type $\gamma$~Doradus stars, driven by convective flux blocking \citep{Guzik+2000} and Slowly Pulsating B-type (SPB) stars, driven by the $\kappa/\gamma$-mechanism \citep{Waelkens+1985, Waelkens+1991, Cox+1992, Pamyatnykh+1999}.
% These pulsators include $\gamma$ Doradus stars driven by the modulation of the radiative flux by convection at the base of a deep envelope convection zone \citep{Guzik+2000}, and Slowly Pulsating B (SPB) stars and $\beta$ Ceph stars driven by the $\kappa$-mechanism \citep{Waelkens+1985, Waelkens+1991, Kiriakidis+1992, Moskalik+1992, Cox+1992}.
These $g$-mode pulsators have been used to provide insights into many aspects of stellar structure and evolution, such as the masses of stellar cores \citep{Johnston+2021, Pedersen+2022} internal mixing processes \citep{Pedersen+2018,Michielsen+2021}, and angular momentum transport \citep{Aerts+2019,Ouazzani2020,Salmon2022,Burssens+2023,Mombarg2023,Moyano2024}. Recent work has also suggested that $g$-modes can be used to probe the binary evolutionary history of stellar mergers in the evolution products of low- and intermediate-mass \citep{Rui+2021} and high-mass stars \citep{Bellinger+2023:2023arXiv231100038B}.

These insights into stellar properties are possible due to the intricate dependence of period spectrum of $g$-mode pulsations on the size of the convective core, and the chemical composition gradient and structure outside of the core \citep[e.g.][]{Dziembowski1993,Miglio+2008,Hatta+2023}. Most current works only consider single-star evolution when inferring stellar properties. However, mass transfer can profoundly influence the structure and composition gradients of accreting stars even after thermal re-adjustment, as indicated by numerous studies using 1D stellar evolution codes \citep{Braun+1995,Renzo+2021,Miszuda+2021}. These changes in structure are usually the result of the rejuvenation and growth of the convective nuclear burning core. As the frequencies of stellar pulsations are finely tuned by the internal structure of stars, asteroseismology holds the potential to identify the signature of previous mass transfer in various classes of pulsating stars. Furthermore, assuming single star evolution for a star that has undergone accretion may result in misleading inferences of its stellar properties from asteroseismology.

Earlier works have explored the asteroseismic modelling of stars in post mass-transfer binaries, with various degrees of accounting for the history of mass transfer. In particular, \citet{Guo2017b, Guo+2017:2017ApJ...837..114G} and \citet{Chen2021} performed asteroseismic analysis of pulsating stars in post-mass transfer binaries observed by {\it Kepler}. While \citet{Guo2017b,Guo+2017:2017ApJ...837..114G} found suitable solutions using only single star evolution models, \citet{Chen2021} compared solutions from binary and single star evolution models. Although \citet{Guo+2017:2017ApJ...837..114G} briefly mention that their solutions may be impacted by not considering the different composition and $g$-mode cavity resulting from mass transfer, they make no further analysis of these effects. Each of these studies conclude that, for their particular systems, single star models were sufficient. 

\citet{Miszuda+2021,Miszuda+2022:2022MNRAS.514..622M} investigated the instability of $p$-modes in two post mass transfer binaries computed with \texttt{MESA}. In their model, the second star accretes nearly conservatively. At the late stages of accretion, the deeper helium-rich layers of the donor star are transferred. The resulting accretor becomes heavily enriched in helium in its outermost layers. They find that this affects overall structure, the mode excitation and pulsation frequencies. In their models, the enriched layer of helium stays is only present at the surface leading to an inversion of average particle mass (mean molecular weight, see their Figure~10). One would expect that mixing processes, such as the Rayleigh-Taylor instability, thermohaline mixing and also rotational mixing \citep{Kippenhahn+1980, Cantiello+2007}, would mix the helium-rich material with the layers below. Both models and observations suggest this overabundance of helium-rich material is erased even before the end of Roche-Lobe overflow \citep{Renzo+2021}.

In this work, we study the longer-lived changes to the internal structure of a rejuvenated accretor.  We investigate the impact on the asteroseismic signals of $g$-modes in accretor star.  We focus on late B-type main sequence stars, which have a relatively high binary fraction and are commonly observed to pulsate as SPB stars. 

We model accretion in a binary system using \texttt{MESA} (\S\ref{sec:methods}) and demonstrate the difference in evolution and internal structure between an accreting star and an equivalent single star, even when accreting non-enriched material (\S\ref{sec:bse}). Using the \texttt{GYRE} stellar oscillation code we then show how this influences the period spacing pattern of the accreting star (\S\ref{sec:asteroseismic}). We highlight how the properties of the star can be inferred inaccurately if single stellar models are used (\S\ref{sec:fitting}). All code to reproduce the results and figures in this paper is available on GitHub\footnote{\url{https://github.com/TomWagg/mass-gainer-seismology}} and Zenodo\footnote{\url{https://zenodo.org/records/10011675}\label{zenodo_note}}. Interactive versions of several figures are available online\footnote{\url{https://www.tomwagg.com/html/interact/mass-gainer-asteroseismology.html}}.

\section{Model and numerical setup} \label{sec:methods}

In this section we outline the setup of our \texttt{MESA} binary model and specify the numeric setup of our \texttt{GYRE} calculations.

\subsection{\texttt{MESA} model setup}\label{sec:model_setup}

We use Modules for Experiments in Stellar Astrophysics \citep[\texttt{MESA},][]{Paxton2011, Paxton2013, Paxton2015, Paxton2018, Paxton2019, Jermyn2023} version r23.05.1 \citep{mesa_zenodo} to simulate non-rotating models for a binary system, as well as a grid of single stars against which to compare. Our full inlists, template folders and our model outputs are available on Zenodo$^{\ref{zenodo_note}}$.

In particular, the most pertinent settings that we use for this work are as follows: We adopt the \citet{Ledoux+1947} criterion to account for the presence of a chemical gradient when determining the stability of convection. We include semiconvection following \citet{Langer+1983} with a scaled efficiency of $\alpha_{\rm sc}=0.1$. We use exponential core overshooting from \cite{Herwig+2000}, setting $(f, f_0) = (0.01, 0.005)$ \citep{Claret+2017}. We set a minimum diffusive mixing coefficient of $20\,{\rm cm^2 \, s^{-1}}$; this smooths out any numerical discontinuities in the composition gradients and partially accounts for the lack of rotational mixing in our models. We motivate our choice of $20\,{\rm cm^2 \, s^{-1}}$ and highlight the effect of changing the mixing coefficient on our results in Appendix~\ref{app:min_D_mix}. We do not account for any rotation in our models. For more details on the input physics and settings, see Appendix~\ref{app:mesa_inputs}. 

Our binary model has a donor with an initial mass of 4$\,{\rm M_\odot}$ and an accretor with an initial mass of 3$\,{\rm M_\odot}$, such that after accretion we form a star in the typical mass range of SPB stars \citep{Waelkens+1985, Waelkens+1991, Kurtz+2022}. We chose an orbital period of 5 days such that the donor will fill its Roche-lobe shortly after leaving the main sequence and undergo so-called Case B mass transfer. We account for non-conservative mass transfer with a mass transfer efficiency of $\beta = 0.5$. We allow the secondary to accrete 0.5$\,{\rm M_\odot}$, which corresponds to about 30\% of the mass that the donor star loses.  The remaining mass is lost from the system taking away approximately the specific angular momentum similar to that of the orbit of the accreting star. We only allow accretion of the outer most layers of the donor star, which are not yet enriched in helium. After the mass transfer phase ends and the donor retreats within it Roche-lobe, we further evolve the accretor until central hydrogen depletion. 

For comparison, we also evolve a grid of single stars with masses from 2-6$\,{\rm M_\odot}$ until the end of helium core burning, using the same physical assumptions. All our models are calculated for a metallicity of $Z = 0.02$.  For a demonstration of the numerical convergence with spatial and temporal resolution of our models, see Appendix~\ref{app:convergence_tests}. See also  Section~\ref{sec:caveats} for a discussion of the model limitations and caveats.

%We evolve this system through Roche-lobe overflow, and proceed with mass transfer until the secondary accretes 0.5$\,{\rm M_\odot}$ of material from the donor. We end mass transfer at this point as we expect the spin-up of the accretor in a rotating model would likely hinder accretion in a similar manner. Although our choice of $0.5$\,{\rm M_\odot}$$ is somewhat arbitrary, the same amount could be obtained through different choices of mass transfer efficiencies. The benefit of this approach is that we avoid accreting helium-rich material from the deeper layers of the donor, which we would not expect for a spun-up accretor.
%At this point we end mass transfer, detach the binary and reduce the donor to a point mass, since we are no longer interested in the properties of the donor and thus only track the evolution of the accretor from this point onwards. By detaching the binary we avoid numerical complications with the late evolutionary phases of the donor which are not of interest to this study, and additionally avoid the regime where rotation would become dominant.
% Although our choice of $0.5$\,{\rm M_\odot}$$ is somewhat arbitrary, this amount could be obtained through different choices of mass transfer efficiencies and the spin-up of the accretor would likely hinder accretion in a similar manner.
%By constructing our model in this way we are able to account for a realistic, variable accretion rate as predicted by the evolution of the orbit and donor star. 

\subsection{\texttt{GYRE} setup} \label{sec: gyre_setup}

We use the \texttt{GYRE} stellar oscillation code \citep[version 7.1, ][]{Townsend+2013,Townsend+2018,Goldstein+2020,Sun+2023} to calculate the periods of the ${\ell = 1}, {m = 0}$ $g$-modes for both our accreting star and equivalent single star models. We calculate the adiabatic eigenfrequencies for dipole ($\ell=1$) modes for each model in our \texttt{MESA} grid and scan 2000 frequency bins on an inversely sampled grid from $0.25$ to $10$~d$^{-1}$ (equivalent to periods from $0.1$ to $4$ days). We set the outer boundary condition to use a vacuum boundary condition and solve the full 6th order dimensionless stellar oscillation equations \citep{Dziembowski+1971, Christensen-Dalsgaard+2008} using the Colloc scheme \texttt{MAGNUS\_GL6}. We use the same \texttt{GYRE} setup for our single and binary star models. Although excitation physics are an interesting avenue to study in future work, we do not consider non-adiabatic calculations in this work.

\section{Binary Stellar Evolution}\label{sec:bse}

In this section, we describe the evolution of our model system, both across the Hertzsprung-Russell diagram and in terms of its internal structure.

\subsection{Hertzsprung-Russell diagram evolution}\label{sec:hrd}

\begin{figure}
    \centering
    \includegraphics[width=\columnwidth]{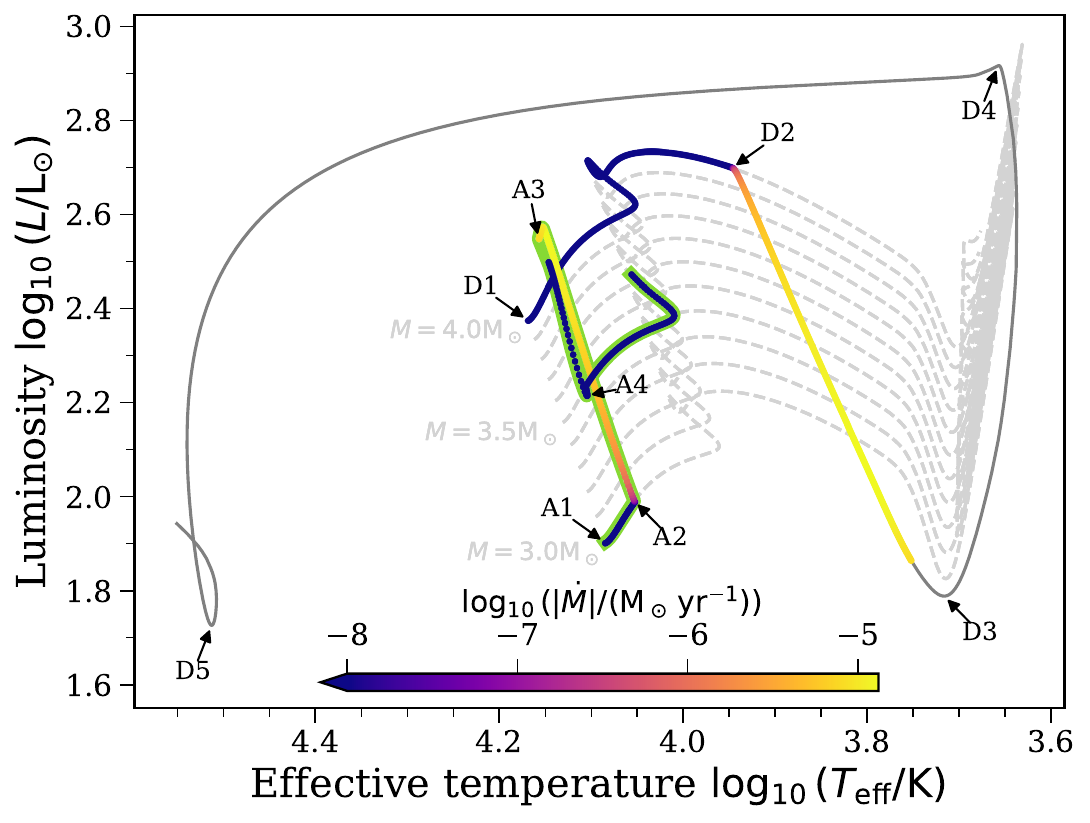}
    \caption{Hertzsprung-Russell diagram showing the evolution of the binary. Tracks are coloured by the mass loss and mass accretion rate for the donor (marked D1-5) and accretor (marked A1-4), see Section~\ref{sec:hrd} for an explanation of the labels. We limit the accretion on the companion to $0.5 \, {\rm M_\odot}$ and follow its further evolution until core hydrogen exhaustion, see Section~\ref{sec:model_setup} for details on the technical implementation. For reference, we show single star tracks as dashed light grey curves, with masses from $3$ to $4 \, {\rm M_\odot}$ in $0.1 \, {\rm M_\odot}$ intervals.\\Interactive plot available online \href{www.tomwagg.com/html/interact/mass-gainer-asteroseismology.html\#fig1}{\faChartArea}.}
    \label{fig:hrd}
\end{figure}

In Figure~\ref{fig:hrd} we show the evolution across the Hertzsprung-Russell diagram of both the donor and accretor of our binary model, with a subset of our single stellar models in the background. Below we explain these tracks, but we note that binary evolution of this nature has been described in many classic papers \citep[e.g.][]{Morton+1960:1960ApJ...132..146M, Smak+1962:1962AcA....12...28S, Paczynski+1966:1966AcA....16..231P,Kippenhahn+1969:1969A&A.....3...83K,Yungelson+1973:1973NInfo..27...93Y,vanderLinden+1987:1987A&A...178..170V} and more recent works \citep[e.g.][]{Yoon+2010:2010ApJ...725..940Y,Claeys+2011:2011A&A...528A.131C, Eldridge+2013:2013MNRAS.436..774E, Tauris+2015:2015MNRAS.451.2123T, McClelland+2016:2016MNRAS.459.1505M, Yoon+2017:2017ApJ...840...10Y, Gotberg+2017:2017A&A...608A..11G, Renzo+2021}.

The evolution of the donor (starting at D1) initially follows the 4$\,{\rm M_\odot}$ single star track, expanding across the main sequence, exhausting core hydrogen and moving across the Hertzsprung gap. During the expansion on the Hertzsprung gap, at point D2, the donor overflows its Roche-lobe and diverges from the single star track. As it loses mass, it is driven out of thermal equilibrium. It decreases in luminosity as a fraction of the photons produced deep inside are now used to do work to expand the outer layers.  The orbit shrinks slightly at first, but quickly starts to widen \citep{Renzo+2019:2019A&A...624A..66R}. At point D3 the donor star is still transferring mass but it starts to regain thermal equilibrium. At point D4 the donor has lost its entire hydrogen-rich envelope and contracts within its Roche-lobe. It keeps contracting until it ignites helium at the point marked D5, where it resides a compact subdwarf.

%As the mass transfer proceeds the orbit of the binary shrinks, increasing the mass transfer rate until the point of minimum Roche-Lobe for the donor (at D3). 
%Beyond this point we show the evolution of the donor with a grey line. 
%With the reversal of the mass ratio, mass transfer causes the orbit to widen, decreasing the mass transfer rate and allowing the donor to increase in luminosity once more. Once the orbit widens to such an extent that Roche-lobe overflow ceases (at point D4), the donor relaxes on a thermal timescale, before proceeding with helium core burning from point D5 onwards.

The evolution of the accretor (starting at A1) follows the 3$\,{\rm M_\odot}$ single star track initially, but early into its main sequence evolution it starts to accrete mass from the donor (at point A2).  This drives the accretor out of thermal equilibrium causing it to expand and increase in luminosity. Once we cease the mass transfer (at point A3) the accretor returns to thermal equilibrium at point A4 and proceeds with its main sequence evolution. At this stage it closely resembles the evolution of a 3.5$\,{\rm M_\odot}$ single star.

\subsection{Rejuvenation and chemical gradients}\label{sec:xh_profiles}

\begin{figure}[tb]
    \centering
    \includegraphics[width=\columnwidth]{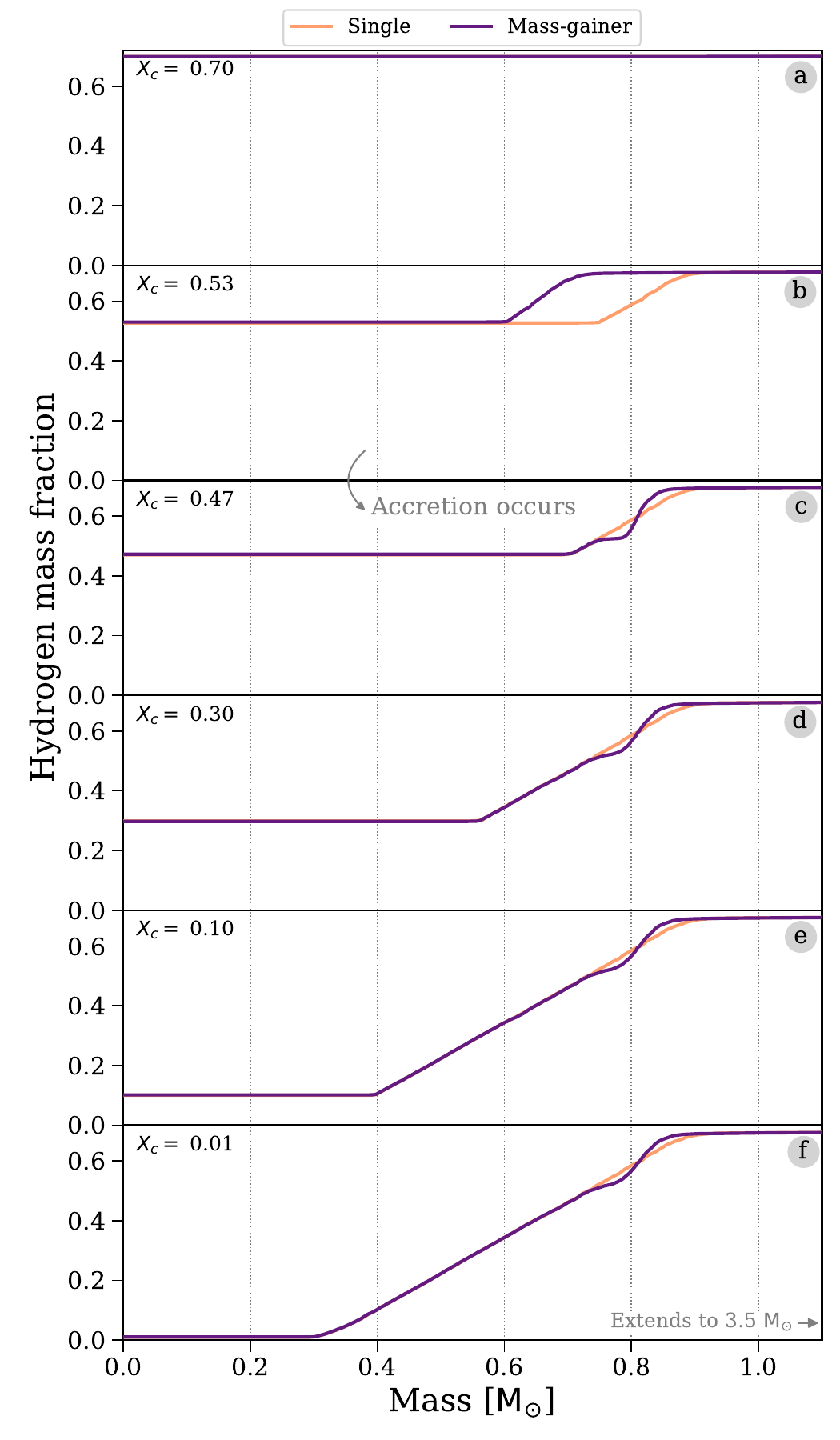}
    \caption{Comparison of the hydrogen abundance profiles between (i) an initially $3 \, {\rm M_\odot}$ star that accretes mass from a companion and (ii) a single star with the same final mass of $3.5 \, {\rm M_\odot}$. Each panel compares the stars at the same central hydrogen abundance, which is annotated in each panel.\\Interactive plot available online \href{www.tomwagg.com/html/interact/mass-gainer-asteroseismology.html\#fig2-4}{\faChartArea}}
    \label{fig:XH_profiles}
\end{figure}

Although the accretor closely follows a more massive single star track in the Hertzsprung-Russell diagram, its \textit{internal} structure has been altered to support the incoming mass after a fraction of its nuclear evolution has already elapsed, leading to enhanced mixing as the convective core expands in mass coordinate \citep{Neo+1977, Hellings1983, Renzo+2023}. This process leaves behind a signature in the hydrogen abundance profile of the star, which we plot in Figure~\ref{fig:XH_profiles}. In each panel we compare the accretor of our binary model with a 3.5$\,{\rm M_\odot}$ single star, thus the stars have the same final mass. Mass transfer occurs between panels b and c and we discuss the differences below.

First, we consider the evolution of the convective stellar core and the abundance profile for the single star. As the star evolves, it burns hydrogen in its core, decreasing the central hydrogen abundance. The reduced hydrogen abundance decreases the opacity of the core, allowing radiation to travel more freely, leading to a recession of the convective core in mass coordinate \citep{Mitalas+1972,Crowe+1982,Miglio+2008,SilvaAguirre+2011, Xin+2022}. As the core recedes it has a decreasing hydrogen abundance, and therefore it imprints a composition gradient in its wake in the abundance profile. We see these trends in Figure~\ref{fig:XH_profiles} as the single star (shown in orange) evolves.

\begin{figure}[tb]
    \centering
    \includegraphics[width=\columnwidth]{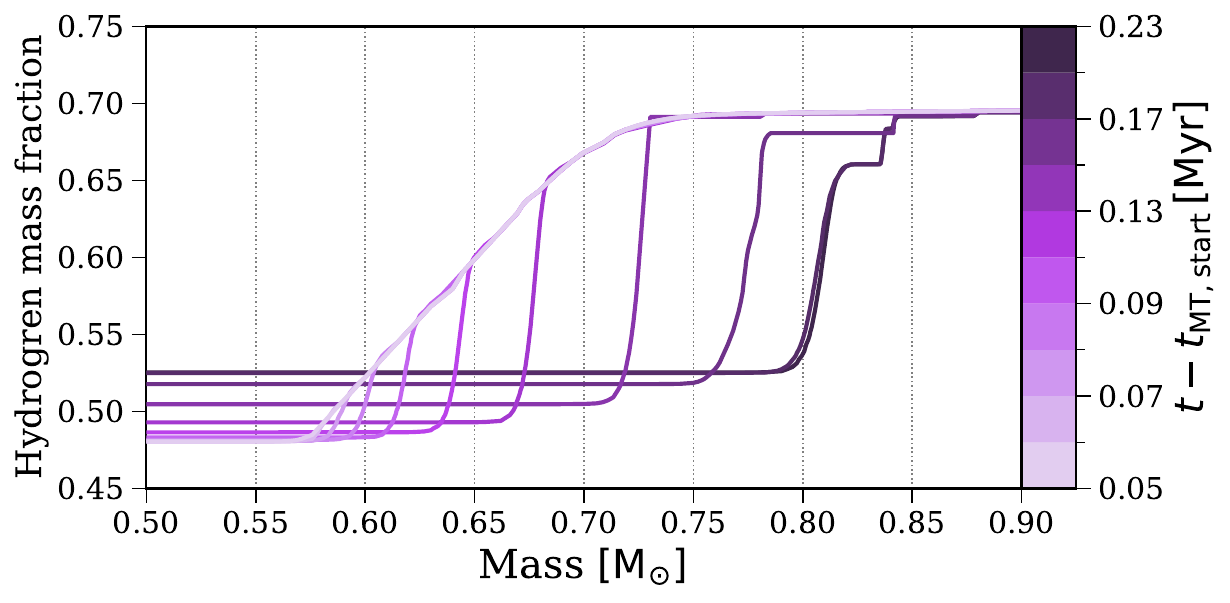}
    \caption{Hydrogen abundance profile of our accretor model during mass transfer. Each line is coloured by its time after the start of mass transfer. This plot shows more time-resolved evolution between panels b and c of Figure~\ref{fig:XH_profiles}.}
    \label{fig:X_H_zoom_MT}
\end{figure}

For the mass-gainer the evolution initially proceeds in a similar manner. In panel \ref{fig:XH_profiles}b, the shape is similar to that of the single star, though with a smaller convective core due to the star's initially lower mass. Between panels \ref{fig:XH_profiles}b and \ref{fig:XH_profiles}c, mass transfer occurs. As mass transfer proceeds the accretor increases in luminosity to compensate for the additional mass. This leads to an increase in the convective core size, which one can see as the profiles move outwards in mass coordinates in Figure~\ref{fig:X_H_zoom_MT}. At the same time, this expansion of the core leads to enhanced convective boundary mixing and a rejuvenation of the accretor as more hydrogen is mixed into the core, increasing the central abundance \citep{Neo+1977}. The expansion of the core into the region through which it previously receded sharpens the composition gradient, resulting in the `kink' in the abundance profile relative to the single star for the remaining panels of Figure~\ref{fig:XH_profiles}. The origin of this feature is shown in Figure~\ref{fig:X_H_zoom_MT}, where we see the hydrogen abundance increase and extend outwards as the core rejuvenates, thus washing away the previous gradient. We emphasise that this occurs even in the absence of rotation and associated mixing. Returning to Figure~\ref{fig:XH_profiles}, the evolution of the abundance profile after mass transfer proceeds similarly to that of a single star, with subsequent recession of the core and a resulting composition gradient. Critically, however, the feature arising from mass transfer remains throughout the main sequence, albeit marginally smoothed by internal mixing.

\section{Asteroseismic Signals} \label{sec:asteroseismic}

In this Section we demonstrate how the differences in internal structure between the accretor and single star lead to altered asteroseismic signals. We first consider how the Brunt–Väisälä (buoyancy) frequency profile is changed, before showing how this influences the period spacing patterns.

\subsection{Brunt–Väisälä frequency profile}\label{sec:bvf}

\begin{figure*}
    \centering
    \includegraphics[width=\textwidth]{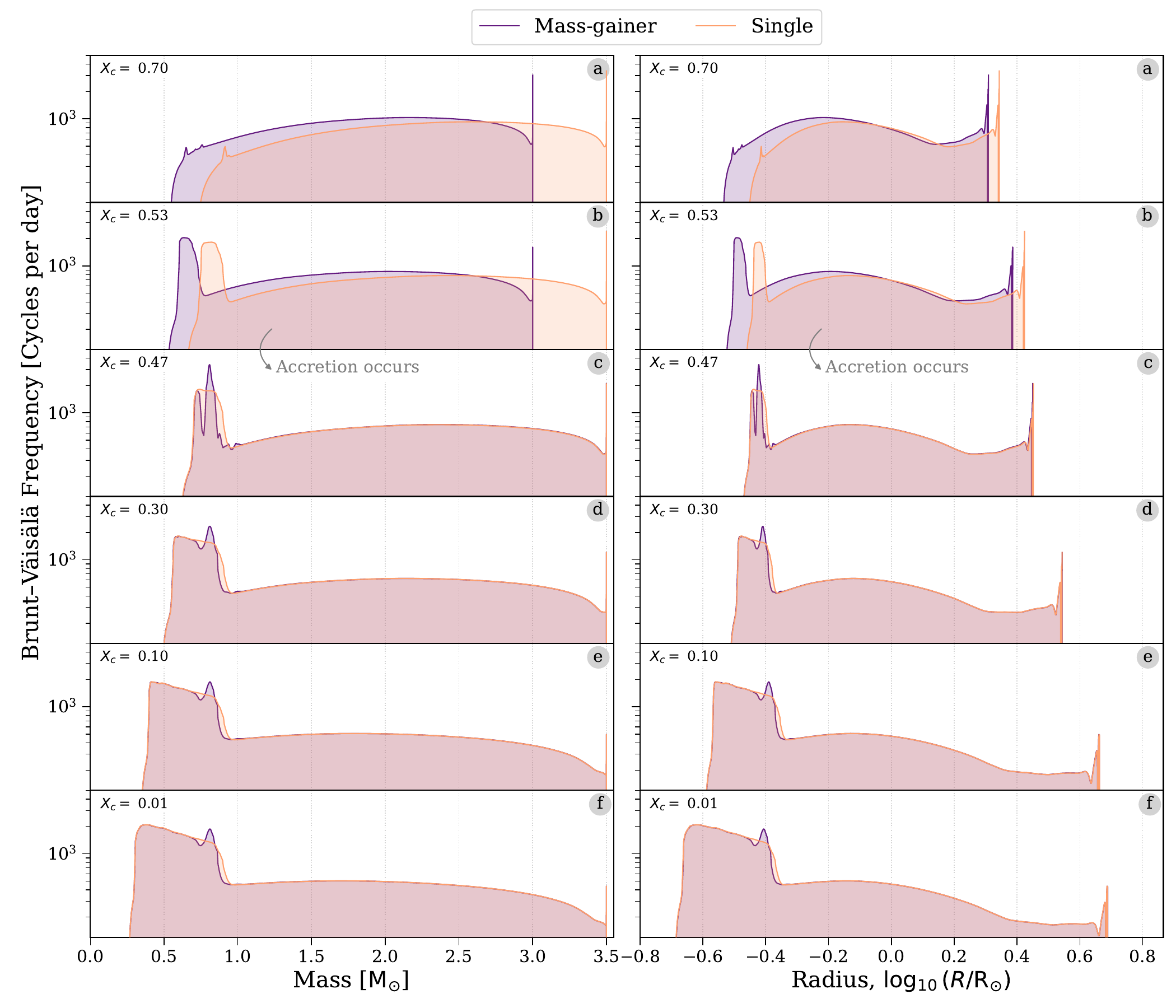}
    \caption{As Figure~\ref{fig:XH_profiles}, but showing the Brunt–Väisälä frequency profile for the same evolutionary timesteps. \textbf{Left:} as a function of mass coordinate, \textbf{right:} as a function of the radial coordinate. Interactive plot available online \href{www.tomwagg.com/html/interact/mass-gainer-asteroseismology.html\#fig2-4}{\faChartArea}}
    \label{fig:BV_profiles}
\end{figure*}

The Brunt–Väisälä frequency \citep{BVF-vaisala, BVF-brunt}, $N$, defines the regions in which convective instabilities can occur, such that $N^2 < 0$ indicates a convective region, and $N^2 > 0$ a radiative region in which $g$-modes can propagate\footnote{The Brunt–Väisälä frequency was originally derived in meteorology and only later applied to stellar evolution. It was first derived in German by \citet{BVF-vaisala} and, despite the typical ordering of the names, independently two years later in English by \citet{BVF-brunt}.}.
Physically, it can be understood as the frequency at which a small element of vertically displaced material will oscillate within a radiative region.
The Brunt–Väisälä frequency directly determines the period distribution of $g$-mode oscillations, and thus it is pertinent to consider the impact of mass transfer on it. For an ideal gas, the frequency can be approximated as
\begin{equation}\label{eq:bvf}
    N^2 \approxeq \frac{g^2 \rho}{P} \qty(\nabla_{\rm ad} - \nabla + \nabla_\mu),
\end{equation}
where $\rho$ is the density, $g$ is the local gravitational acceleration, $P$ is the pressure, $\nabla_{\rm ad} \approx 2/5$ is the adiabatc temperature gradient and assumed to be a constant, $\nabla \equiv \dv{\ln T}{\ln P}$ is the temperature gradient and $\nabla_\mu$ is the chemical composition gradient. Although many of the terms in this expression are similar for our accretor model and equivalent single star model, the density profile and, as noted in Section~\ref{sec:xh_profiles}, the composition gradient $\nabla_\mu$ show significant differences and as such we expect similar differences in the Brunt–Väisälä frequency profile.

In Figure~\ref{fig:BV_profiles} we compare, for the same central hydrogen content, the Brunt–Väisälä frequency profiles for the accretor star model and single star model with the same final mass. Each panel is for the same central hydrogen content as in Figure~\ref{fig:XH_profiles} for a simple comparison. We additionally show the profile as a both a function of mass coordinate and radial coordinate in the two columns.

Considering first the single star model, we see that initially the convective core ($N^2<0$) extends to ${\sim}$0.75$\,{\rm M_\odot}$ (or ${\sim}0.3\,{\rm R_\odot}$) and the frequency profile changes smoothly across the star. As the star evolves, the core recedes, leaving behind a chemical gradient; a peak then emerges in the Brunt–Väisälä frequency profile that extends between the core and the unmixed outer regions of the star. This peak is directly due to the chemical composition gradient ($\nabla_\mu$) imprinted on the star by the receding core during the main sequence. As the star evolves, the peak extends in concert with the recession of the core, in line with the composition gradient.

For the accretor model, we see similar evolution in panels \ref{fig:BV_profiles}a and \ref{fig:BV_profiles}b (before mass transfer occurs). Immediately following mass transfer (in panel \ref{fig:BV_profiles}c), the convective core radial extent and mass coordinate align with the single star model and several distinct features emerge outside of the core, arising due to the kink in the composition gradient visible in panel \ref{fig:XH_profiles}c. As the star evolves, chemical mixing smooths these features to some extent, but importantly the star retains a double-peaked Brunt–Väisälä frequency profile for the rest of its main sequence evolution.

We are unaware of any process in single star evolution that would result in an equivalent Brunt–Väisälä frequency profile for the accretor. In single star evolution we expect a smooth monotonic change in the chemical composition gradient due to the recession of the convective core. Therefore, the profile would always have a smooth, unimodal peak. The occurrence of an increase in convective core size in a rejuvenated accretor results in a change in the chemical composition gradient not possible in single stars \citep{Renzo+2023}.

\subsection{Period spacing patterns}\label{sec:period_spacing}

All differences between the mass-gainer and equivalent single star that we have noted so far are within the internal structure, and so are not directly observable. Therefore, we now consider the impact of these internal structure changes on the observable period spacing pattern.

The period spacing pattern is defined as the difference in period between modes of the same spherical degree, $\ell$, and neighbouring radial order, $n$. Under the assumption of spherical symmetry and high radial order ($n \gg \ell$), this difference is constant and follows the asymptotic $g$-mode period spacing given by \citet{Tassoul+1980}:
\begin{equation}\label{eq:period-spacing}
    \Delta P_g = \frac{\pi^2}{\sqrt{\ell(\ell+1)}} \qty[\int_{r_0}^{r_1} \frac{N}{r} \dd{r}]^{-1},
\end{equation}
where $\ell$ is the spherical degree, $N$ is the Brunt–Väisälä frequency (see Eq.~\ref{eq:bvf}) and $r_0$ and $r_1$ are the boundaries of the $g$-mode oscillation cavity, which in our model correspond to the convective core boundary and the outer edge of the star respectively.

Deviations from the asymptotic period spacing occur due to abrupt shifts in the Brunt–Väisälä frequency profile, which trap particular modes in certain regions of the star, altering their periods relative to the regular pattern \citep[e.g.][]{Dziembowski1993,Miglio+2008}. The sensitivity of these deviations to the Brunt–Väisälä frequency thus makes the period spacing pattern a useful observable for probing the internal structure of a star \citep[e.g.][]{Aerts+2010}.

\begin{figure}%[bt]
    \centering
    \includegraphics[width=\columnwidth]{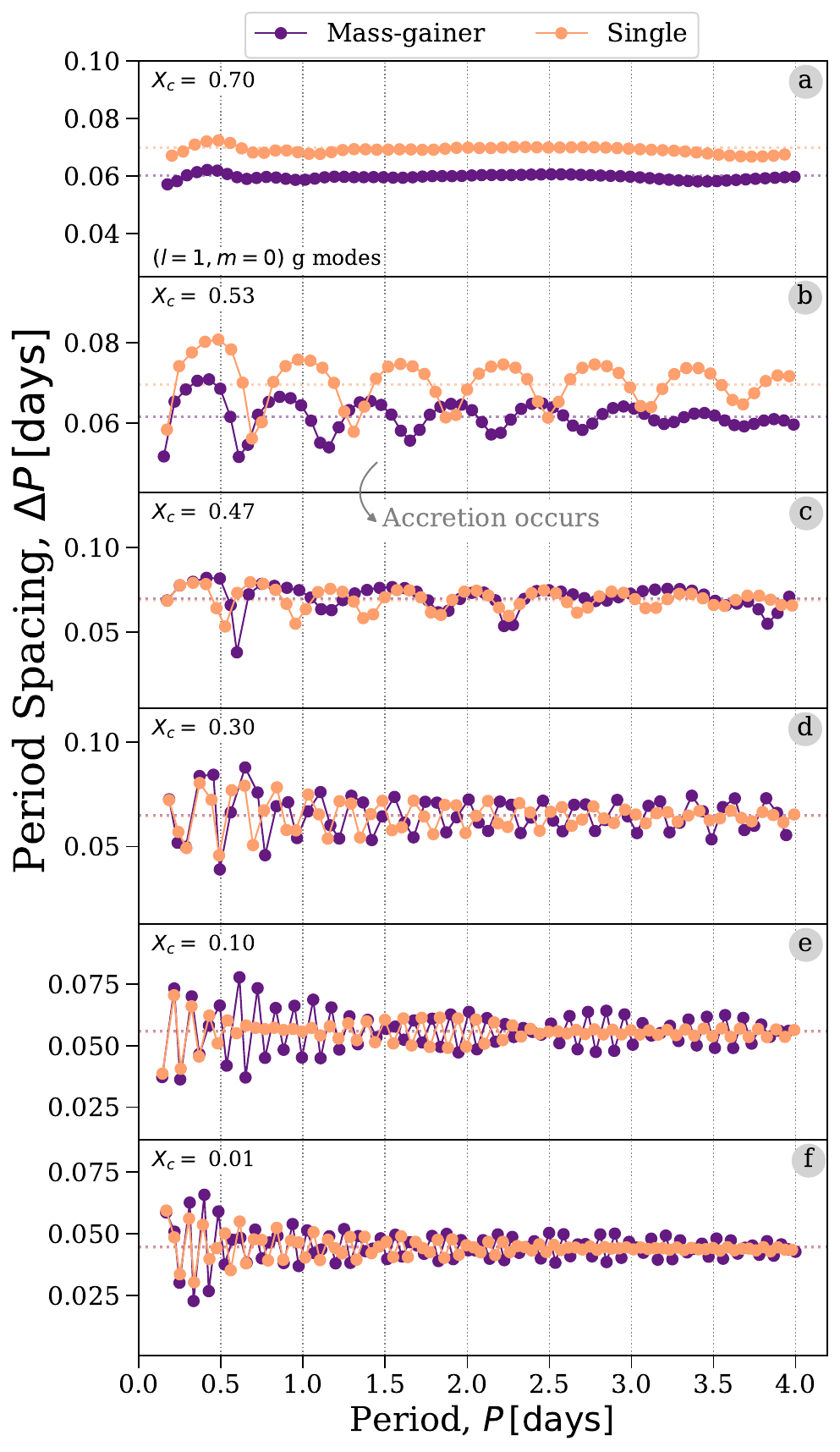}
    \caption{As Figure~\ref{fig:XH_profiles}, but showing the period spacing patterns of the $\ell = 1, m = 0$ $g$-modes. Asymptotic period spacings for each model are shown as dotted lines. The $y$-axis limits vary by panel. Interactive plot available online \href{www.tomwagg.com/html/interact/mass-gainer-asteroseismology.html\#fig5}{\faChartArea}}
    \label{fig:period_spacing}
\end{figure}

In Figure~\ref{fig:period_spacing} we compare the period spacing pattern of a mass-gainer to that of an equivalent single star at different stages during their evolution. At the zero-age main sequence (in panel \ref{fig:period_spacing}a), the period spacing pattern closely follows each star's asymptotic period spacing (denoted as dotted lines) due to the lack of any composition gradients. Early during the main sequence, immediately prior to mass transfer (in panel \ref{fig:period_spacing}b), the pattern now displays some oscillation around the asymptotic value due to the chemical composition gradients that have developed outside of the core. Since this is currently pre-accretion, the stars have different masses and thus convective core sizes, resulting in an offset between their asymptotic period spacings.

In subsequent panels (\ref{fig:period_spacing}c--f) there are several differences in the period spacing pattern, despite the fact that the stars now have the same mass and convective core size. The two main differences can be expressed in terms of the amplitude and phase of oscillations in the period spacing pattern. Frequently in the stars' later evolution, the amplitude of deviations from the asymptotic spacing are larger for the mass-gainer (for example between ${\sim}0.5$--$1.0$ days in panel \ref{fig:period_spacing}e). This is because the mass-gainer contains regions with steeper chemical composition gradients \citep{Renzo+2023}, which more strongly impact the Brunt–Väisälä frequency and thus the period of oscillations.

% \help{It seems like the amplitude argument could be flawed, if this was true would we not only see amplitude differences in regions with phase differences?}

% \tom{From Earl: on the other hand, all the modes touch essentially all parts of the BV profile, so all modes are sensitive at some level to this change}

In addition, we find that the oscillations in the period spacing pattern shift phase in certain regions for the mass gainer. This is most apparent in panel e, in which the patterns are out of phase for periods between ${\sim}1.5$--$3.2 \, {\rm days}$ and in-phase otherwise. These period-dependent shifts arise due to difference between the Brunt–Väisälä frequency profiles occurring in the region of changing chemical composition. Certain modes are more sensitive to certain regions of the star than others. Modes that are more sensitive to the region of changing chemical composition are shifted and so move out of phase, whilst other modes are less sensitive to the differences from a single star and thus oscillate with the same periods, remaining in-phase.

\begin{figure*}[tb]
    \centering
    \includegraphics[width=\textwidth]{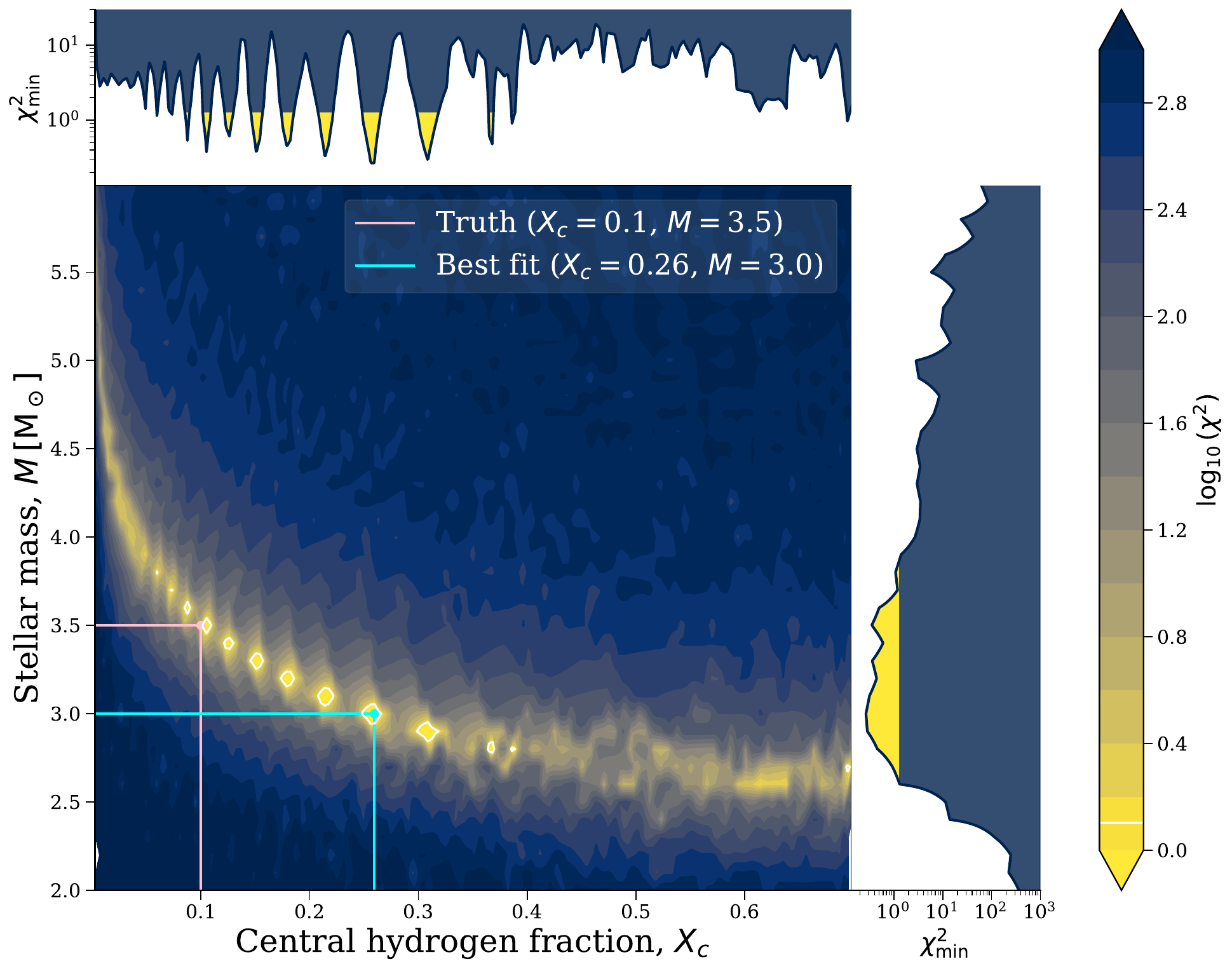}
    \caption{Stellar properties could be inferred incorrectly when assuming single star evolution if one does not consider the full multimodal posteriors. Main panel shows the $\chi^2$ values for fitting our $M = 3.5 \, {\rm M_\odot}$ mass-gainer model at $X_c = 0.1$ with our entire grid of single star models. The true and best fit values are highlighted with lines. We add white contour lines at $\chi^2_{\rm best} + 1$ to highlight models that are statistically compatible with the true value. Side panels show marginal distributions (the minimum $\chi^2$ marginalised over a given axis), where any values with $\chi^2_{\rm min} \le \chi^2_{\rm best} + 1$ are filled with yellow to mark them as statistically compatible.}
    \label{fig:chi2}
\end{figure*}

\section{Fitting accretors with single stars} \label{sec:fitting}

We have demonstrated that, compared to an equivalent single star, a mass-gainer shows significant differences in its period spacing pattern as a result of accretion altering its chemical composition gradient. Therefore, given that around $20\%$ SPB stars are expected to be in interacting binaries \citep{Sana+2012,deMink+2014}, modelling that assumes single star evolution may result in incorrect inferences.

We test how incorrect these inferences may be by fitting the period spacing pattern of our mass-gainer model assuming single star evolution. We use \texttt{GYRE} to compute the periods of $\ell = 1, m = 0$ $g$-modes for our grid of single star models between 2 and 6$\,{\rm M_\odot}$ across the entire main sequence. We then perform a $\chi^2$ fit for a given mass-gainer period spacing pattern with every single star model, at every timestep.

We match the periods of models \textit{independently} of radial order, since, in reality, the exact radial order of an observed pulsation is not known a priori. This means that the best fitting period for a given radial order in the mass-gainer model may actually be from a different radial order in the single star model. We therefore need to determine the optimal matching between these sets of periods. We make the assumption that the `observed' period spacing pattern is continuous and monotonic in radial order. This simplifies the matching process, as for each single star model we must now only determine the offset in radial order, $\epsilon$, from the mass-gainer model, such that a period $P_{{\rm mg}, i}$ in the mass-gainer corresponds to a period $P_{{\rm s}, (i + \epsilon)}$ in the single star.

In summary, we calculate the $\chi^2$ for each model as
\begin{equation}
    \chi^2 = \sum_{i}^{N} \frac{(P_{{\rm mg}, i} - P_{{\rm s}, (i + \epsilon)})^2}{\sigma_i^2},
\end{equation}
where $P_{\rm mg}$ and $P_{\rm s}$ are the $g$-mode periods of the mass-gainer and single star models respectively, and $\sigma_i$ is the uncertainty on the measurement. For the purposes of this investigation we adopt a frequency uncertainty of $1 / 1150 \, {\rm days}$ based on the Kepler time baseline.

In Figure~\ref{fig:chi2}, we show an example of this $\chi^2$ fitting. In this plot the mass-gainer model has a mass of $3.5 \,{\rm M_\odot}$ and central hydrogen content of $X_c = 0.1$. %The pit-like features in the $\chi^2$ values occur each time the offset $\epsilon$ is shifted by one during the fitting \citep[e.g.][]{Buysschaert+2018:2018A&A...616A.148B}.
The pit-like features in the $\chi^2$ values are the result of degeneracies in the underlying models. Specifically, as the core mass decreases with decreasing $X_c$, the average $\Delta P$ values decrease, thus creating a degeneracy between higher mass models with lower $X_c$ and lower mass models with high $X_c$ \citep[e.g., ][]{Buysschaert+2018:2018A&A...616A.148B, Mombarg2019}.

The best fitting model when assuming single star evolution underestimates the mass at $3.0 \,{\rm M_\odot}$ and overestimates the central hydrogen content at $X_c = 0.26$, more than twice the true value. However, this best-fitting value is found along the trough of degeneracy between mass and hydrogen content. Each value within the white contour lines, which includes the true value, is each statistically compatible with the mass-gainer.

Therefore, when fitting a potential mass-gainer with single star models, it is critical to consider the full multimodal posterior distributions. For example, \citealt{Basu+2012:2012ApJ...746...76B} suggests selecting all models with likelihoods over
95\% of the value of the best-fitting model. This allows one to account for the large degeneracies in models. We highlight that if one were to simply take the mean and standard deviation of the mode of the posterior, then one would infer inaccurate properties for the star (Figure~\ref{fig:chi2}).

\begin{figure}
    \centering
    \includegraphics[width=\columnwidth]{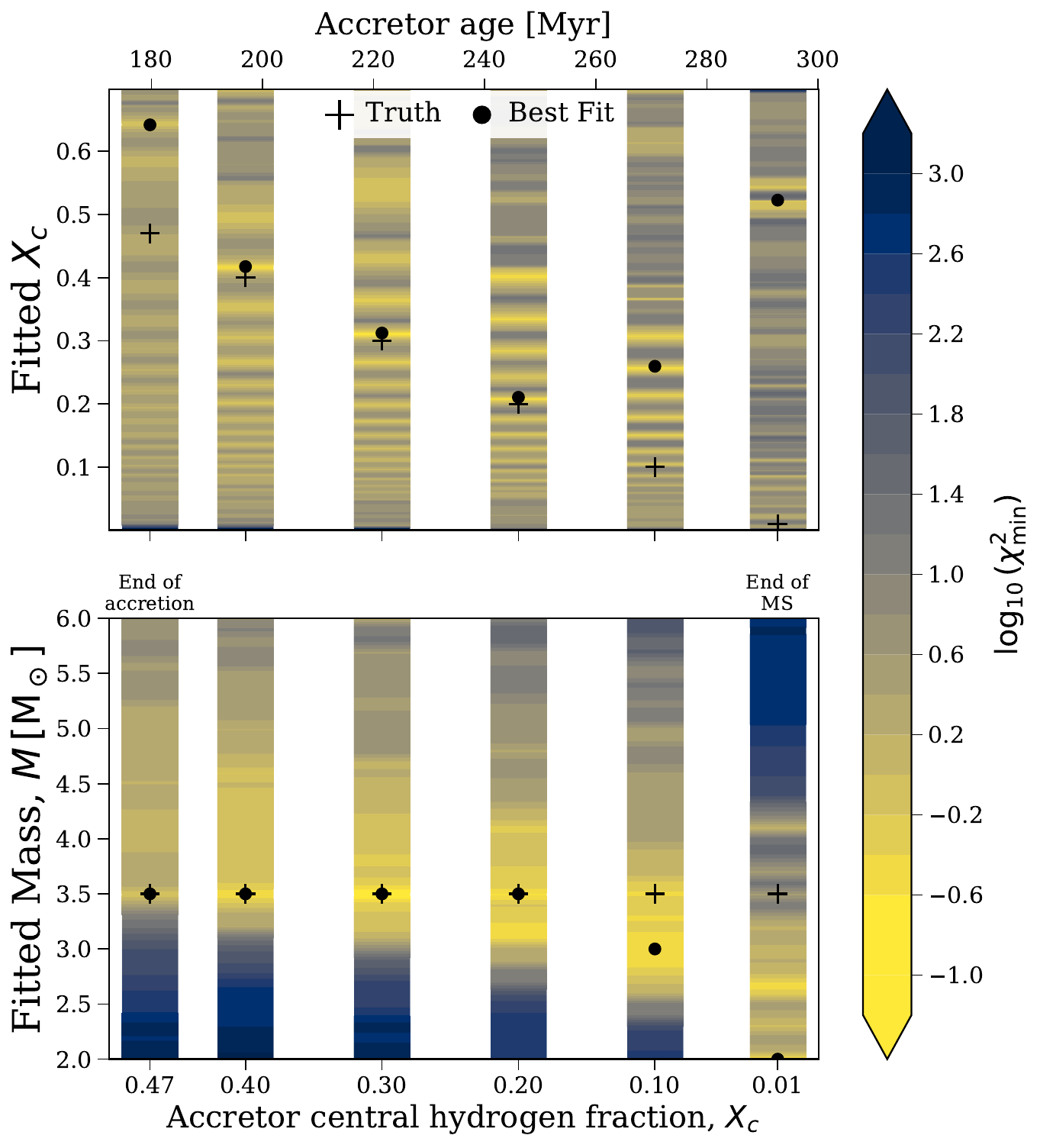}
    \caption{Assuming single star evolution for stars that have accreted mass can lead to an overestimation of central hydrogen fraction, $X_c$, and an underestimation of mass when not accounting for full multimodal posteriors. Comparison between fitted single star models and the `true' values of the mass-gainer model for $X_c$ (upper panel) and mass (lower panel) as a function of $X_c$ from the end of accretion until the end of the main sequence. Each column shows, for a given $X_c$, a contour plot with the minimum $\chi^2$ values (similar to Figure~\ref{fig:chi2}) annotated with markers for the true value and absolute best fit.}
    \label{fig:relative_estimates_psp}
\end{figure}

We repeated this fitting procedure throughout the post-accretion evolution of the mass-gainer model, as shown in Figure~\ref{fig:relative_estimates_psp}. Each column in this Figure shows the same information as Figure~\ref{fig:chi2} at a different evolutionary stage. At evolutionary stages recently after accretion has ceased (larger values, close to $X_c = 0.47$), the best-fit values of the mass and $X_c$ from the single star modes are considered statistically compatible with the `true' values from the accretor model. However, as the main sequence proceeds ($X_c < 0.1$) the lowest $\chi^2$ model diverges from the true value, tending to overestimate $X_c$ and underestimate $M$. Furthermore, we highlight that the range of the parameter space that is consistent with observations is generally large, and shows multiple optima. This stresses the importance of considering the full posterior distributions.

\subsection{Stellar ages}

\begin{figure}
    \centering
    \includegraphics[width=\columnwidth]{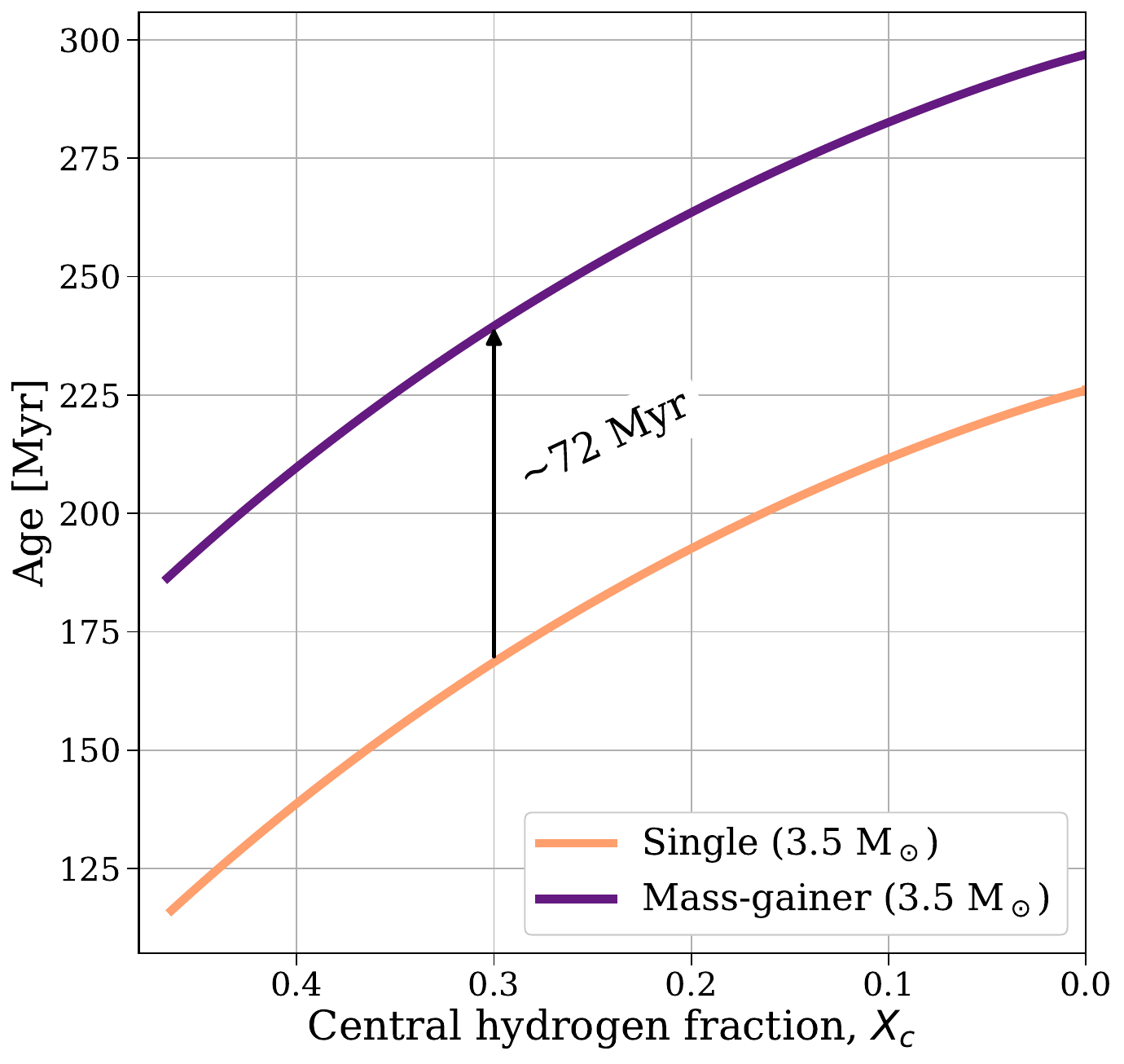}
    \caption{A single star model with the same age and $X_c$ as a mass-gainer will underestimate the age. Comparison of the stellar age of our mass-gainer model and $3.5 \, {\rm M_\odot}$ single star model during post-accretion evolution.}
    \label{fig:age-Xc}
\end{figure}

Even when stellar mass and $X_c$ are well fit by single star models, inferences of stellar ages for mass-gainers will be incorrect. A mass-gainer is initially less massive and thus evolves more slowly before accretion occurs. For this reason, our mass-gainer model is ${\sim}72 \, {\rm Myr}$ older than the equivalent ($3.5 \, {\rm M_\odot}$) single star model for all values of $X_c$ post-accretion (Figure~\ref{fig:age-Xc}). The difference in age is dependent on how much mass is accreted and at what stage during the main sequence it is accreted. A star that accreted more mass, or accreted it later in its main sequence, would have a more significant difference in age. Overall, using asteroseismology and single star models to infer the age of mass-gainers will consistently underestimate stellar ages.

\subsection{Improved uncertainties}

For our assumed frequency uncertainty (based on the Kepler time baseline), we have shown that a single star model with the same stellar mass and $X_c$ is statistically compatible with our mass-gainer model. However, with only moderate improvements (${\sim}15\%$, see Figure~\ref{fig:chi2}) to the uncertainties one can rule out this single star model to $2\sigma$.

Therefore, as the quality of data improves with future detectors we find that mass-gainers will be increasingly distinguishable from single star models with the correct mass and $X_c$. This provides an opportunity to better study the effects of mass transfer and accretion. Yet at the same time, this highlights the increasing importance in considering a star's accretion history when using asteroseismology to infer its stellar properties.

\section{Discussion}
%\subsection{Caveats}
\label{sec:caveats}

\paragraph{Binary parameters and the treatment of mass transfer} The mass-gainer model that we use was allowed to accrete only 0.5$\,{\rm M_\odot}$ of material. This was partially motivated by the idea that an accretor with rotation is expected to quickly reach critical rotation and prevent further accretion \citep[e.g.][]{Petrovic+2005}, others do argue that accretion is possible even beyond critical rotation \citep{Popham+1991, Paczynski+1991}. The amount of accreted material will likely affect the quantitative results. If the star had accreted significantly more, its new convective core may have further expanded, possibly erasing the chemical composition gradient and thus tracers of mass transfer and its asteroseismic signals. If the star had accreted significantly less, the imprints on the structure would be milder and probably harder to detect.  

Another parameter is the evolutionary state of the secondary star at the onset of mass transfer (which is governed by the initial mass ratio and, to a lesser extent, the initial period of the binary system). In our model, the accretion occurred when the accretor was still relatively unevolved (at a central hydrogen mass fraction of 0.53). If the secondary had been more evolved, it would had a more pronounced internal chemical gradient, which would have been harder to fully erase. This is not only because the gradient then covers a larger range in mass coordinate, but also since the presence of a steep gradient may have an inhibiting effect to mixing \citep{Braun+1995}. 

Variations on the amount of accreted material and the evolutionary stage of the secondary should be explored systematically in future work.

%In this work, we do not vary the amount of mass transferred, or the evolutionary state of the accretor at the onset of mass transfer. Both of these parameters are likely to impact the imprint left on the accretors internal structure, and thus its asteroseismic signals. 

%, though it well motivated since an accretor with rotation is expected to quickly reach critical rotation and prevent further accretion \citep[e.g.][]{Petrovic+2005}, others do argue that accretion is possible even beyond critical rotation \citep{Popham+1991, Paczynski+1991}.

%Additionally, we only consider one mass-gainer model for this proof of principle analysis and, though we expect the differences we find will be prevalent across a range of masses and periods, further work is necessary to confirm this.

\paragraph{Rotation} We have limited the scope of our investigations by neglecting rotation in each model. For slow to moderate rotating $g$-mode pulsators, many recent studies only add the effects of rotation at the stage of calculating pulsation frequencies \citep[e.g.][]{Michielsen+2021}. Yet, mass-gainers could reach close-to-critical rotation during mass transfer due to the exchange of angular momentum \citep{Packet+1981, deMink+2013:2013ApJ...764..166D, Renzo+2021}, and this could impact several aspects of our results \citep[e.g.][and references therein]{Aerts+2023:2023arXiv231108453A}. In practice, such high rotation rates are not found in observations and accretors typically have spins that are only 10-40\% of the critical rate (\citealp{Dervisoglu+2010}, see however e.g., \citealp{Zehe+2018:2018AN....339...46Z}), but this may still significantly impact the asteroseismic signatures.

The two dominant effects concern rotational mixing and the shifting of $g$-modes due to the introduction of the Coriolis force. Rotation is expected to enhance internal chemical mixing throughout the star through mechanisms such as shear mixing, dynamical instabilities, and Eddington-Sweet circulation \citep{Maeder2000}. These mechanisms are expected to act in the vicinity of the core as well and thus alter the chemical composition gradient, which is the key difference between the mass-gainer and single star models. By introducing the Coriolis force, rotation actively shifts the periods of $g$-modes \citep{Townsend2003}, introducing a characteristic tilt in the period spacing pattern \citep{Bouabid+2013}. In addition to the effects of modified chemical mixing profile, rotation will further complicate the construction of observed period spacing patterns. However, this does open up the opportunity to investigate the core rotation of stars that have undergone mass transfer, as was done by \citet{Guo2019}. Future work should investigate the role of rotation in modifying these predictions.

\paragraph{Treatment of steep chemical gradients} As a result of mass accretion and rejuvenation, accretors develop a steep chemical composition gradient. The treatment of this in a 1D evolutionary code is uncertain and numerically challenging \citep[e.g.\ ][]{Lau+2014:2014A&A...570A.125L}. We enforce a minimum level of diffusive mixing in our models that smooths the chemical composition gradient. This has an impact on our results for the Brunt–Väisälä frequency and period spacing pattern. However, as we demonstrate in Appendix~\ref{app:min_D_mix}, differences between the mass-gainer and single star models are still present for a wide-range of choices for this parameter. Crucially, we note that this may artificially suppress physical signatures in the model that result from rejuvenation. Since mass transfer is a rapid and ill-understood phenomenon, we cannot say for certain whether or not all of the signatures that it imparts on the stellar structure are numerical or physical. Specifically, we refer to the points marked as `glitches' in the Brunt–Väisälä profile in Figure~\ref{fig:min_D_mix}, which are preferentially smoothed as \texttt{min\_D\_mix} increases. We further note that increased \texttt{min\_D\_mix} values can serve as a proxy for the effects of slow to moderate rotation on chemical mixing in the stellar interior. Thus, while mass transfer may impart some small scale features in the chemical profile of the accretor, we can expect them to be suppressed over time by chemical mixing induced by slow to moderate rotation.

\section{Summary \& Conclusions} \label{sec:conclusion}

We have presented new simulations investigating the impact of mass transfer on the asteroseismic signals of slowly pulsating B stars. We used \texttt{MESA} to evolve a binary star system and computed the asteroseismic properties of the accretor star using \texttt{GYRE}. We compared the internal structure, and rejuvenation, of the accretor, as well as its period spacing pattern, to an equivalent single star. Our main conclusions are as follows:

\begin{enumerate}
    \setlength\itemsep{1em}
    \item \textbf{Mass transfer produces a distinct asteroseismic imprint}\\Our accretor star model shows a significantly different internal chemical composition gradient \citep[e.g.][]{Renzo+2021}, and hence Brunt–Väisälä frequency\ (see Figures~\ref{fig:XH_profiles}--\ref{fig:BV_profiles}), even though we only consider non-enriched hydrogen being accreted. While \citet{Miszuda+2021} demonstrated that mass transfer impacts the frequencies of $p$ modes in lower mass $\delta$ Scuti type stars, they considered helium enriched materials being accreted. Even when considering only un-enriched material, the modified chemical gradient selectively traps certain $g$-modes and therefore produces a measurably different asteroseismic signal (see Figure~\ref{fig:period_spacing}) compared to an equal mass star that has undergone single star evolution.
    \item {\bf The asteroseismic signature of rejuvenation persists throughout the main-sequence evolution of the mass gainer}\\ At times soon after rejuvenation, there is only a small region of the star with a non-zero chemical composition gradient (e.g.\ Figure~\ref{fig:XH_profiles}c). Therefore, fewer modes are shifted by the steep chemical gradient. As the star evolves and the region with a non-zero chemical gradient increases, more modes are trapped by the gradient, and even though mixing modifies the remaining gradient, it is present until the terminal-age main sequence. As such, we can see that the differences in the period-spacing patterns of the mass-gainer and single star persist until the end of the main sequence as well. In particular, we notice that the mass gainer has increased oscillatory amplitude in the period-spacing pattern compared to the single star, likely due to the steeper gradient feature.
    % Hence, modes which are trapped in the accretor model show significant differences in their period-spacing patterns when compared to patterns from models that have undergone single star evolution.
    \item {\bf Single-star models cannot robustly reproduce period spacing patterns from mass accreting models}\\We demonstrate that, given realistic observational uncertainties on pulsation period-spacing patterns, we cannot uniquely identify the matching accretor model when using a single-star evolution model (Figure~\ref{fig:chi2}). While the individual maximum a-posteriori point estimates correctly identifies the stellar mass and $X_c$ in many cases, the range of models that are statistically valid covers a much wider parameter range in all cases (Figure~\ref{fig:relative_estimates_psp}). Moreover, even models that correctly estimate mass and $X_c$ significantly underestimate the stellar age (Figure~\ref{fig:age-Xc}).

    \item {\bf The asteroseismic signature of rejuvenation can be identified with detailed modelling} \\
    While we have demonstrated that single star models are able to recover the bulk properties of mass gaining models, there is still clear structure in the period-spacing patterns that can be modelled. As accretor stars are necessarily in binary systems, this provides a unique opportunity to leverage the high-precision fundamental stellar parameter estimates $(<1\%)$ in the asteroseismic modelling procedure \citep[e.g.][]{Torres2010,Johnston2019a,Sekaran2021}. With an independent and precise estimate of the mass gainer stars mass, radius, and age, the asteroseismic analysis will be driven by small difference in the trapping pattern as opposed to matching the asymptotic pattern value \citep{Johnston2019b}. Fortuitously, there are numerous examples of stars that have undergone mass transfer in eclipsing binaries at all masses. 
    % \item \textbf{It is critical to consider full posterior distributions when fitting a mass-gainer}\\We show that, if one considers only the best fitting model, one may infer inaccurate stellar properties when fitting the period spacing pattern of a star that has undergone accretion (see Figure~\ref{fig:chi2}). For our mass-gainer model we find that, though the best-fitting model can be inaccurate, a single star model with the same values of $X_c$ and $M$ remains statistically compatible throughout the post-accretion evolution.
\end{enumerate}
This work demonstrates that asteroseismology can be used to probe the structural impact of mass accretion in binary interaction. These results have immediate implications as this opens the door to providing observational constraints on poorly-understood binary evolution processes that are otherwise unobservable. In particular, there are several classes of known or strongly favoured post-mass transfer systems, such as Algol variables \citep[e.g. ][]{Shi2022} and Be-type stars \citep[e.g.][]{Baade1982,Bodensteiner+2020,LabadieBartz2022}, that are: i) known $g$-mode pulsators, and ii) have existing long time-base photometry with telescopes such as Kepler \citep{Borucki+2010} and TESS \citep{Ricker+2015}. Future work will investigate the sensitivity of asteroseismic signals to the rate of mass transfer, as well as the mixing processes that accompany mass transfer. Finally, detecting the signatures of mass transfer in currently available asteroseismic data will bridge the exploits of asteroseismology to help calibrate predictions of gravitational wave progenitor populations which require at least one episode of mass transfer in their evolution to interpret the observed distributions of gravitational wave progenitors and events \citep{Abbott+2023,Renzo+2023}.

\begin{acknowledgements}
    We thank the Kavli Foundation and the Max Planck Institute for Astrophysics for supporting the 2023 Kavli Summer Program during which much of this work was completed. In particular, the authors thank Isabel Thapa, Stephen Justham, Mahdieh Schmidt and Pascala Garaud for their incredible efforts in organising this program. TW also thanks UW Subject Library Matthew Parsons for helping to track down the original text of \citet{BVF-vaisala} and Donna Thompson for aiding in listing it on NASA ADS. TW applauds Ruggero Valli for his help, and patience, with setting up \texttt{MESA} on the MPA cluster and sharing the secret commands of \texttt{kinit} and \texttt{aklog}. CJ gratefully acknowledges support from the Netherlands Research School of Astronomy (NOVA). CJ thanks Amadeusz Miszuda for sharing his {\sc mesa} {\sc binary} inlists to help start up this research. RT acknowledges support from NASA grant 80NSSC20K0515.
\end{acknowledgements}

\textit{Software:} \texttt{MESA} \citep{Paxton2011, Paxton2013, Paxton2015, Paxton2018, Paxton2019, Jermyn2023} version r23.05.01 \citep{mesa_zenodo}, \texttt{GYRE} \citep{Townsend+2013, Townsend+2018}, \texttt{Astropy} \citep{astropy:2013, astropy:2018, astropy:2022}, \texttt{Python} \citep{python}, \texttt{numpy} \citep{numpy}, \texttt{pandas} \citep{pandas_1.4.2, pandas_paper}, \texttt{matplotlib} \citep{matplotlib}, \texttt{scipy} \citep{Virtanen+2020}

\bibliographystyle{aa}
\bibliography{all}

\begin{thebibliography}{149}
\expandafter\ifx\csname natexlab\endcsname\relax\def\natexlab#1{#1}\fi

\bibitem[{{Abbott} {et~al.}(2023){Abbott}, {Abbott}, {Acernese}, {Ackley},
  {Adams}, {Adhikari}, {Adhikari}, {Adya}, {Affeldt}, {Agarwal}, {Agathos},
  {Agatsuma}, {Aggarwal}, {Aguiar}, {Aiello}, {Ain}, {Ajith}, {Akutsu}, {de
  Alarc{\'o}n}, {Akcay}, {Albanesi}, {Allocca}, {Altin}, {Amato}, {Anand},
  {Anand}, {Ananyeva}, {Anderson}, {Anderson}, {Ando}, {Andrade}, {Andres},
  {Andri{\'c}}, {Angelova}, {Ansoldi}, {Antelis}, {Antier}, {Antonini},
  {Appert}, {Arai}, {Arai}, {Arai}, {Araki}, {Araya}, {Araya}, {Areeda},
  {Ar{\`e}ne}, {Aritomi}, {Arnaud}, {Arogeti}, {Aronson}, {Arun}, {Asada},
  {Asali}, {Ashton}, {Aso}, {Assiduo}, {Aston}, {Astone}, {Aubin}, {Austin},
  {Babak}, {Badaracco}, {Bader}, {Badger}, {Bae}, {Bae}, {Baer}, {Bagnasco},
  {Bai}, {Baiotti}, {Baird}, {Bajpai}, {Ball}, {Ballardin}, {Ballmer},
  {Balsamo}, {Baltus}, {Banagiri}, {Bankar}, {Barayoga}, {Barbieri}, {Barish},
  {Barker}, {Barneo}, {Barone}, {Barr}, {Barsotti}, {Barsuglia}, {Barta},
  {Bartlett}, {Barton}, {Bartos}, {Bassiri}, {Basti}, {Bawaj}, {Bayley},
  {Baylor}, {Bazzan}, {B{\'e}csy}, {Bedakihale}, {Bejger}, {Belahcene},
  {Benedetto}, {Beniwal}, {Bennett}, {Bentley}, {Benyaala}, {Bergamin},
  {Berger}, {Bernuzzi}, {Berry}, {Bersanetti}, {Bertolini}, {Betzwieser},
  {Beveridge}, {Bhandare}, {Bhardwaj}, {Bhattacharjee}, {Bhaumik}, {Bilenko},
  {Billingsley}, {Bini}, {Birney}, {Birnholtz}, {Biscans}, {Bischi},
  {Biscoveanu}, {Bisht}, {Biswas}, {Bitossi}, {Bizouard}, {Blackburn}, {Blair},
  {Blair}, {Blair}, {Bobba}, {Bode}, {Boer}, {Bogaert}, {Boldrini}, {Bonavena},
  {Bondu}, {Bonilla}, {Bonnand}, {Booker}, {Boom}, {Bork}, {Boschi}, {Bose},
  {Bose}, {Bossilkov}, {Boudart}, {Bouffanais}, {Bozzi}, {Bradaschia}, {Brady},
  {Bramley}, {Branch}, {Branchesi}, {Brandt}, {Brau}, {Breschi}, {Briant},
  {Briggs}, {Brillet}, {Brinkmann}, {Brockill}, {Brooks}, {Brooks}, {Brown},
  {Brunett}, {Bruno}, {Bruntz}, {Bryant}, {Bulik}, {Bulten}, {Buonanno},
  {Buscicchio}, {Buskulic}, {Buy}, {Byer}, {Cadonati}, {Cagnoli}, {Cahillane},
  {Bustillo}, {Callaghan}, {Callister}, {Calloni}, {Cameron}, {Camp}, {Canepa},
  {Canevarolo}, {Cannavacciuolo}, {Cannon}, {Cao}, {Cao}, {Capocasa}, {Capote},
  {Carapella}, {Carbognani}, {Carlin}, {Carney}, {Carpinelli}, {Carrillo},
  {Carullo}, {Carver}, {Diaz}, {Casentini}, {Castaldi}, {Caudill},
  {Cavagli{\`a}}, {Cavalier}, {Cavalieri}, {Ceasar}, {Cella},
  {Cerd{\'a}-Dur{\'a}n}, {Cesarini}, {Chaibi}, {Chakravarti}, {Subrahmanya},
  {Champion}, {Chan}, {Chan}, {Chan}, {Chan}, {Chan}, {Chandra}, {Chanial},
  {Chao}, {Chapman-Bird}, {Charlton}, {Chase}, {Chassande-Mottin},
  {Chatterjee}, {Chatterjee}, {Chatterjee}, {Chaturvedi}, {Chaty},
  {Chatziioannou}, {Chen}, {Chen}, {Chen}, {Chen}, {Chen}, {Chen}, {Chen},
  {Chen}, {Cheng}, {Cheong}, {Cheung}, {Chia}, {Chiadini}, {Chiang},
  {Chiarini}, {Chierici}, {Chincarini}, {Chiofalo}, {Chiummo}, {Cho}, {Cho},
  {Choudhary}, {Choudhary}, {Christensen}, {Chu}, {Chu}, {Chu}, {Chua},
  {Chung}, {Ciani}, {Ciecielag}, {Cie{\'s}lar}, {Cifaldi}, {Ciobanu}, {Ciolfi},
  {Cipriano}, {Cirone}, {Clara}, {Clark}, {Clark}, {Clarke}, {Clearwater},
  {Clesse}, {Cleva}, {Coccia}, {Codazzo}, {Cohadon}, {Cohen}, {Cohen},
  {Colleoni}, {Collette}, {Colombo}, {Colpi}, {Compton}, {Constancio}, {Conti},
  {Cooper}, {Corban}, {Corbitt}, {Cordero-Carri{\'o}n}, {Corezzi}, {Corley},
  {Cornish}, {Corre}, {Corsi}, {Cortese}, {Costa}, {Cotesta}, {Coughlin},
  {Coulon}, {Countryman}, {Cousins}, {Couvares}, {Coward}, {Cowart}, {Coyne},
  {Coyne}, {Creighton}, {Creighton}, {Criswell}, {Croquette}, {Crowder},
  {Cudell}, {Cullen}, {Cumming}, {Cummings}, {Cunningham}, {Cuoco},
  {Cury{\l}o}, {Dabadie}, {Canton}, {Dall'Osso}, {D{\'a}lya}, {Dana},
  {Daneshgaranbajastani}, {D'Angelo}, {Danila}, {Danilishin}, {D'Antonio},
  {Danzmann}, {Darsow-Fromm}, {Dasgupta}, {Datrier}, {Datta}, {Dattilo},
  {Dave}, {Davier}, {Davies}, {Davis}, {Davis}, {Daw}, {Dean}, {Debra},
  {Deenadayalan}, {Degallaix}, {de Laurentis}, {Del{\'e}glise}, {Del Favero},
  {de Lillo}, {de Lillo}, {Del Pozzo}, {Demarchi}, {de Matteis}, {D'Emilio},
  {Demos}, {Dent}, {Depasse}, {de Pietri}, {De Rosa}, {de Rossi}, {Desalvo},
  {de Simone}, {Dhurandhar}, {D{\'\i}az}, {Diaz-Ortiz}, {Didio}, {Dietrich},
  {di Fiore}, {di Fronzo}, {di Giorgio}, {di Giovanni}, {di Giovanni}, {di
  Girolamo}, {di Lieto}, {Ding}, {di Pace}, {di Palma}, {di Renzo},
  {Divakarla}, {Dmitriev}, {Doctor}, {D'Onofrio}, {Donovan}, {Dooley},
  {Doravari}, {Dorrington}, {Drago}, {Driggers}, {Drori}, {Ducoin}, {Dupej},
  {Durante}, {D'Urso}, {Duverne}, {Dwyer}, {Eassa}, {Easter}, {Ebersold},
  {Eckhardt}, {Eddolls}, {Edelman}, {Edo}, {Edy}, {Effler}, {Eguchi},
  {Eichholz}, {Eikenberry}, {Eisenmann}, {Eisenstein}, {Ejlli}, {Engelby},
  {Enomoto}, {Errico}, {Essick}, {Estell{\'e}s}, {Estevez}, {Etienne}, {Etzel},
  {Evans}, {Evans}, {Ewing}, {Fafone}, {Fair}, {Fairhurst}, {Farah}, {Farinon},
  {Farr}, {Farr}, {Farrow}, {Fauchon-Jones}, {Favaro}, {Favata}, {Fays},
  {Fazio}, {Feicht}, {Fejer}, {Fenyvesi}, {Ferguson}, {Fernandez-Galiana},
  {Ferrante}, {Ferreira}, {Fidecaro}, {Figura}, {Fiori}, {Fishbach}, {Fisher},
  {Fittipaldi}, {Fiumara}, {Flaminio}, {Floden}, {Fong}, {Font}, {Fornal},
  {Forsyth}, {Franke}, {Frasca}, {Frasconi}, {Frederick}, {Freed}, {Frei},
  {Freise}, {Frey}, {Fritschel}, {Frolov}, {Fronz{\'e}}, {Fujii}, {Fujikawa},
  {Fukunaga}, {Fukushima}, {Fulda}, {Fyffe}, {Gabbard}, {Gadre}, {Gair},
  {Gais}, {Galaudage}, {Gamba}, {Ganapathy}, {Ganguly}, {Gao}, {Gaonkar},
  {Garaventa}, {Garc{\'\i}a}, {Garc{\'\i}a-N{\'u}{\~n}ez},
  {Garc{\'\i}a-Quir{\'o}s}, {Garufi}, {Gateley}, {Gaudio}, {Gayathri}, {Ge},
  {Gemme}, {Gennai}, {George}, {George}, {Gerberding}, {Gergely}, {Gewecke},
  {Ghonge}, {Ghosh}, {Ghosh}, {Ghosh}, {Ghosh}, {Giacomazzo}, {Giacoppo},
  {Giaime}, {Giardina}, {Gibson}, {Gier}, {Giesler}, {Giri}, {Gissi},
  {Glanzer}, {Gleckl}, {Godwin}, {Golomb}, {Goetz}, {Goetz}, {Gohlke},
  {Goncharov}, {Gonz{\'a}lez}, {Gopakumar}, {Gosselin}, {Gouaty}, {Gould},
  {Grace}, {Grado}, {Granata}, {Granata}, {Grant}, {Gras}, {Grassia}, {Gray},
  {Gray}, {Greco}, {Green}, {Green}, {Gretarsson}, {Gretarsson}, {Griffith},
  {Griffiths}, {Griggs}, {Grignani}, {Grimaldi}, {Grimm}, {Grote}, {Grunewald},
  {Gruning}, {Guerra}, {Guidi}, {Guimaraes}, {Guix{\'e}}, {Gulati}, {Guo},
  {Guo}, {Gupta}, {Gupta}, {Gupta}, {Gustafson}, {Gustafson}, {Guzman}, {Ha},
  {Haegel}, {Hagiwara}, {Haino}, {Halim}, {Hall}, {Hamilton}, {Hammond}, {Han},
  {Haney}, {Hanks}, {Hanna}, {Hannam}, {Hannuksela}, {Hansen}, {Hansen},
  {Hanson}, {Harder}, {Hardwick}, {Haris}, {Harms}, {Harry}, {Harry},
  {Hartwig}, {Hasegawa}, {Haskell}, {Hasskew}, {Haster}, {Hattori}, {Haughian},
  {Hayakawa}, {Hayama}, {Hayes}, {Healy}, {Heidmann}, {Heidt}, {Heintze},
  {Heinze}, {Heinzel}, {Heitmann}, {Hellman}, {Hello}, {Helmling-Cornell},
  {Hemming}, {Hendry}, {Heng}, {Hennes}, {Hennig}, {Hennig}, {Hernandez},
  {Vivanco}, {Heurs}, {Hild}, {Hill}, {Himemoto}, {Hines}, {Hiranuma},
  {Hirata}, {Hirose}, {Hochheim}, {Hofman}, {Hohmann}, {Holcomb}, {Holland},
  {Hollows}, {Holmes}, {Holt}, {Holz}, {Hong}, {Hopkins}, {Hough}, {Hourihane},
  {Howell}, {Hoy}, {Hoyland}, {Hreibi}, {Hsieh}, {Hsu}, {Huang}, {Huang},
  {Huang}, {Huang}, {Huang}, {Huang}, {H{\"u}bner}, {Huddart}, {Hughey}, {Hui},
  {Hui}, {Husa}, {Huttner}, {Huxford}, {Huynh-Dinh}, {Ide}, {Idzkowski},
  {Iess}, {Ikenoue}, {Imam}, {Inayoshi}, {Ingram}, {Inoue}, {Ioka}, {Isi},
  {Isleif}, {Ito}, {Itoh}, {Iyer}, {Izumi}, {Jaberianhamedan}, {Jacqmin},
  {Jadhav}, {Jadhav}, {James}, {Jan}, {Jani}, {Janquart}, {Janssens},
  {Janthalur}, {Jaranowski}, {Jariwala}, {Jaume}, {Jenkins}, {Jenner}, {Jeon},
  {Jeunon}, {Jia}, {Jin}, {Johns}, {Jones}, {Jones}, {Jones}, {Jones}, {Jones},
  {Jonker}, {Ju}, {Jung}, {Jung}, {Junker}, {Juste}, {Kaihotsu}, {Kajita},
  {Kakizaki}, {Kalaghatgi}, {Kalogera}, {Kamai}, {Kamiizumi}, {Kanda},
  {Kandhasamy}, {Kang}, {Kanner}, {Kao}, {Kapadia}, {Kapasi}, {Karat},
  {Karathanasis}, {Karki}, {Kashyap}, {Kasprzack}, {Kastaun}, {Katsanevas},
  {Katsavounidis}, {Katzman}, {Kaur}, {Kawabe}, {Kawaguchi}, {Kawai},
  {Kawasaki}, {K{\'e}f{\'e}lian}, {Keitel}, {Key}, {Khadka}, {Khalili}, {Khan},
  {Khazanov}, {Khetan}, {Khursheed}, {Kijbunchoo}, {Kim}, {Kim}, {Kim}, {Kim},
  {Kim}, {Kim}, {Kimball}, {Kimura}, {Kinley-Hanlon}, {Kirchhoff}, {Kissel},
  {Kita}, {Kitazawa}, {Kleybolte}, {Klimenko}, {Knee}, {Knowles}, {Knyazev},
  {Koch}, {Koekoek}, {Kojima}, {Kokeyama}, {Koley}, {Kolitsidou}, {Kolstein},
  {Komori}, {Kondrashov}, {Kong}, {Kontos}, {Koper}, {Korobko}, {Kotake},
  {Kovalam}, {Kozak}, {Kozakai}, {Kozu}, {Kringel}, {Krishnendu}, {Kr{\'o}lak},
  {Kuehn}, {Kuei}, {Kuijer}, {Kulkarni}, {Kumar}, {Kumar}, {Kumar}, {Kumar},
  {Kume}, {Kuns}, {Kuo}, {Kuo}, {Kuromiya}, {Kuroyanagi}, {Kusayanagi},
  {Kuwahara}, {Kwak}, {Lagabbe}, {Laghi}, {Lalande}, {Lam}, {Lamberts},
  {Landry}, {Landry}, {Lane}, {Lang}, {Lange}, {Lantz}, {La Rosa},
  {Lartaux-Vollard}, {Lasky}, {Laxen}, {Lazzarini}, {Lazzaro}, {Leaci},
  {Leavey}, {Lecoeuche}, {Lee}, {Lee}, {Lee}, {Lee}, {Lee}, {Lee}, {Lehmann},
  {Lema{\^\i}tre}, {Leonardi}, {Leroy}, {Letendre}, {Levesque}, {Levin},
  {Leviton}, {Leyde}, {Li}, {Li}, {Li}, {Li}, {Li}, {Li}, {Lin}, {Lin}, {Lin},
  {Lin}, {Lin}, {Linde}, {Linker}, {Linley}, {Littenberg}, {Liu}, {Liu}, {Liu},
  {Liu}, {Llamas}, {Llorens-Monteagudo}, {Lo}, {Lockwood}, {Loh}, {London},
  {Longo}, {Lopez}, {Portilla}, {Lorenzini}, {Loriette}, {Lormand}, {Losurdo},
  {Lott}, {Lough}, {Lousto}, {Lovelace}, {Lucaccioni}, {L{\"u}ck}, {Lumaca},
  {Lundgren}, {Luo}, {Lynam}, {Macas}, {Macinnis}, {MacLeod}, {MacMillan},
  {Macquet}, {Hernandez}, {Magazz{\`u}}, {Magee}, {Maggiore}, {Magnozzi},
  {Mahesh}, {Majorana}, {Makarem}, {Maksimovic}, {Maliakal}, {Malik}, {Man},
  {Mandic}, {Mangano}, {Mango}, {Mansell}, {Manske}, {Mantovani}, {Mapelli},
  {Marchesoni}, {Marchio}, {Marion}, {Mark}, {M{\'a}rka}, {M{\'a}rka},
  {Markakis}, {Markosyan}, {Markowitz}, {Maros}, {Marquina}, {Marsat},
  {Martelli}, {Martin}, {Martin}, {Martinez}, {Martinez}, {Martinez},
  {Martinovic}, {Martynov}, {Marx}, {Masalehdan}, {Mason}, {Massera},
  {Masserot}, {Massinger}, {Masso-Reid}, {Mastrogiovanni}, {Matas},
  {Mateu-Lucena}, {Matichard}, {Matiushechkina}, {Mavalvala}, {McCann},
  {McCarthy}, {McClelland}, {McClincy}, {McCormick}, {McCuller}, {McGhee},
  {McGuire}, {McIsaac}, {McIver}, {McRae}, {McWilliams}, {Meacher}, {Mehmet},
  {Mehta}, {Meijer}, {Melatos}, {Melchor}, {Mendell}, {Menendez-Vazquez},
  {Menoni}, {Mercer}, {Mereni}, {Merfeld}, {Merilh}, {Merritt}, {Merzougui},
  {Meshkov}, {Messenger}, {Messick}, {Meyers}, {Meylahn}, {Mhaske}, {Miani},
  {Miao}, {Michaloliakos}, {Michel}, {Michimura}, {Middleton}, {Milano},
  {Miller}, {Miller}, {Miller}, {Miller}, {Millhouse}, {Mills}, {Milotti},
  {Minazzoli}, {Minenkov}, {Mio}, {Mir}, {Miravet-Ten{\'e}s}, {Mishra},
  {Mishra}, {Mistry}, {Mitra}, {Mitrofanov}, {Mitselmakher}, {Mittleman},
  {Miyakawa}, {Miyamoto}, {Miyazaki}, {Miyo}, {Miyoki}, {Mo}, {Modafferi},
  {Moguel}, {Mogushi}, {Mohapatra}, {Mohite}, {Molina}, {Molina-Ruiz},
  {Mondin}, {Montani}, {Moore}, {Moraru}, {Morawski}, {More}, {Moreno},
  {Moreno}, {Mori}, {Morisaki}, {Moriwaki}, {Morr{\'a}s}, {Mours}, {Mow-Lowry},
  {Mozzon}, {Muciaccia}, {Mukherjee}, {Mukherjee}, {Mukherjee}, {Mukherjee},
  {Mukherjee}, {Mukund}, {Mullavey}, {Munch}, {Mu{\~n}iz}, {Murray},
  {Musenich}, {Muusse}, {Nadji}, {Nagano}, {Nagano}, {Nagar}, {Nakamura},
  {Nakano}, {Nakano}, {Nakashima}, {Nakayama}, {Napolano}, {Nardecchia},
  {Narikawa}, {Naticchioni}, {Nayak}, {Nayak}, {Negishi}, {Neil}, {Neilson},
  {Nelemans}, {Nelson}, {Nery}, {Neubauer}, {Neunzert}, {Ng}, {Ng}, {Nguyen},
  {Nguyen}, {Nguyen}, {Quynh}, {Ni}, {Nichols}, {Nishizawa}, {Nissanke},
  {Nitoglia}, {Nocera}, {Norman}, {North}, {Nozaki}, {Siles}, {Nuttall},
  {Oberling}, {O'Brien}, {Obuchi}, {O'Dell}, {Oelker}, {Ogaki}, {Oganesyan},
  {Oh}, {Oh}, {Oh}, {Ohashi}, {Ohishi}, {Ohkawa}, {Ohme}, {Ohta}, {Okada},
  {Okutani}, {Okutomi}, {Olivetto}, {Oohara}, {Ooi}, {Oram}, {O'Reilly},
  {Ormiston}, {Ormsby}, {Ortega}, {O'Shaughnessy}, {O'Shea}, {Oshino},
  {Ossokine}, {Osthelder}, {Otabe}, {Ottaway}, {Overmier}, {Pace}, {Pagano},
  {Page}, {Pagliaroli}, {Pai}, {Pai}, {Palamos}, {Palashov}, {Palomba}, {Pan},
  {Pan}, {Panda}, {Pang}, {Pang}, {Pankow}, {Pannarale}, {Pant}, {Panther},
  {Paoletti}, {Paoli}, {Paolone}, {Parisi}, {Park}, {Park}, {Parker},
  {Pascucci}, {Pasqualetti}, {Passaquieti}, {Passuello}, {Patel}, {Pathak},
  {Patricelli}, {Patron}, {Paul}, {Payne}, {Pedraza}, {Pegoraro}, {Pele},
  {Arellano}, {Penn}, {Perego}, {Pereira}, {Pereira}, {Perez}, {P{\'e}rigois},
  {Perkins}, {Perreca}, {Perri{\`e}s}, {Petermann}, {Petterson}, {Pfeiffer},
  {Pham}, {Phukon}, {Piccinni}, {Pichot}, {Piendibene}, {Piergiovanni},
  {Pierini}, {Pierro}, {Pillant}, {Pillas}, {Pilo}, {Pinard}, {Pinto}, {Pinto},
  {Piotrzkowski}, {Piotrzkowski}, {Pirello}, {Pitkin}, {Placidi}, {Planas},
  {Plastino}, {Pluchar}, {Poggiani}, {Polini}, {Pong}, {Ponrathnam},
  {Popolizio}, {Porter}, {Poulton}, {Powell}, {Pracchia}, {Pradier},
  {Prajapati}, {Prasai}, {Prasanna}, {Pratten}, {Principe}, {Prodi},
  {Prokhorov}, {Prosposito}, {Prudenzi}, {Puecher}, {Punturo}, {Puosi},
  {Puppo}, {P{\"u}rrer}, {Qi}, {Quetschke}, {Quitzow-James}, {Raab},
  {Raaijmakers}, {Radkins}, {Radulesco}, {Raffai}, {Rail}, {Raja}, {Rajan},
  {Ramirez}, {Ramirez}, {Ramos-Buades}, {Rana}, {Rapagnani}, {Rapol}, {Ray},
  {Raymond}, {Raza}, {Razzano}, {Read}, {Rees}, {Regimbau}, {Rei}, {Reid},
  {Reid}, {Reitze}, {Relton}, {Renzini}, {Rettegno}, {Reza}, {Rezac}, {Ricci},
  {Richards}, {Richardson}, {Richardson}, {Riemenschneider}, {Riles},
  {Rinaldi}, {Rink}, {Rizzo}, {Robertson}, {Robie}, {Robinet}, {Rocchi},
  {Rodriguez}, {Rolland}, {Rollins}, {Romanelli}, {Romano}, {Romel},
  {Romero-Rodr{\'\i}guez}, {Romero-Shaw}, {Romie}, {Ronchini}, {Rosa}, {Rose},
  {Rosi{\'n}ska}, {Ross}, {Rowan}, {Rowlinson}, {Roy}, {Roy}, {Roy}, {Rozza},
  {Ruggi}, {Ryan}, {Sachdev}, {Sadecki}, {Sadiq}, {Sago}, {Saito}, {Saito},
  {Sakai}, {Sakai}, {Sakellariadou}, {Sakuno}, {Salafia}, {Salconi}, {Saleem},
  {Salemi}, {Samajdar}, {Sanchez}, {Sanchez}, {Sanchez}, {Sanchis-Gual},
  {Sanders}, {Sanuy}, {Saravanan}, {Sarin}, {Sassolas}, {Satari},
  {Sathyaprakash}, {Sato}, {Sato}, {Sauter}, {Savage}, {Sawada}, {Sawant},
  {Sawant}, {Sayah}, {Schaetzl}, {Scheel}, {Scheuer}, {Schiworski}, {Schmidt},
  {Schmidt}, {Schnabel}, {Schneewind}, {Schofield}, {Sch{\"o}nbeck}, {Schulte},
  {Schutz}, {Schwartz}, {Scott}, {Scott}, {Seglar-Arroyo}, {Sekiguchi},
  {Sekiguchi}, {Sellers}, {Sengupta}, {Sentenac}, {Seo}, {Sequino}, {Sergeev},
  {Setyawati}, {Shaffer}, {Shahriar}, {Shams}, {Shao}, {Sharma}, {Sharma},
  {Shawhan}, {Shcheblanov}, {Shibagaki}, {Shikauchi}, {Shimizu}, {Shimoda},
  {Shimode}, {Shinkai}, {Shishido}, {Shoda}, {Shoemaker}, {Shoemaker},
  {Shyamsundar}, {Sieniawska}, {Sigg}, {Singer}, {Singh}, {Singh}, {Singha},
  {Sintes}, {Sipala}, {Skliris}, {Slagmolen}, {Slaven-Blair}, {Smetana},
  {Smith}, {Smith}, {Soldateschi}, {Somala}, {Somiya}, {Son}, {Soni}, {Soni},
  {Sordini}, {Sorrentino}, {Sorrentino}, {Sotani}, {Soulard}, {Souradeep},
  {Sowell}, {Spagnuolo}, {Spencer}, {Spera}, {Srinivasan}, {Srivastava},
  {Srivastava}, {Staats}, {Stachie}, {Steer}, {Steinhoff}, {Steinlechner},
  {Steinlechner}, {Stevenson}, {Stops}, {Stover}, {Strain}, {Strang},
  {Stratta}, {Strunk}, {Sturani}, {Stuver}, {Sudhagar}, {Sudhir}, {Sugimoto},
  {Suh}, {Sullivan}, {Summerscales}, {Sun}, {Sun}, {Sunil}, {Sur}, {Suresh},
  {Sutton}, {Suzuki}, {Suzuki}, {Swinkels}, {Szczepa{\'n}czyk}, {Szewczyk},
  {Tacca}, {Tagoshi}, {Tait}, {Takahashi}, {Takahashi}, {Takamori}, {Takano},
  {Takeda}, {Takeda}, {Talbot}, {Talbot}, {Tanaka}, {Tanaka}, {Tanaka},
  {Tanaka}, {Tanaka}, {Tanasijczuk}, {Tanioka}, {Tanner}, {Tao}, {Tao},
  {Mart{\'\i}n}, {Taranto}, {Tasson}, {Telada}, {Tenorio}, {Terhune},
  {Terkowski}, {Thirugnanasambandam}, {Thomas}, {Thomas}, {Thomas}, {Thompson},
  {Thondapu}, {Thorne}, {Thrane}, {Tiwari}, {Tiwari}, {Tiwari}, {Toivonen},
  {Toland}, {Tolley}, {Tomaru}, {Tomigami}, {Tomura}, {Tonelli},
  {Torres-Forn{\'e}}, {Torrie}, {E Melo}, {T{\"o}yr{\"a}}, {Trapananti},
  {Travasso}, {Traylor}, {Trevor}, {Tringali}, {Tripathee}, {Troiano},
  {Trovato}, {Trozzo}, {Trudeau}, {Tsai}, {Tsai}, {Tsang}, {Tsang}, {Tsao},
  {Tse}, {Tso}, {Tsubono}, {Tsuchida}, {Tsukada}, {Tsuna}, {Tsutsui},
  {Tsuzuki}, {Turbang}, {Turconi}, {Tuyenbayev}, {Ubhi}, {Uchikata},
  {Uchiyama}, {Udall}, {Ueda}, {Uehara}, {Ueno}, {Ueshima}, {Unnikrishnan},
  {Uraguchi}, {Urban}, {Ushiba}, {Utina}, {Vahlbruch}, {Vajente}, {Vajpeyi},
  {Valdes}, {Valentini}, {Valsan}, {van Bakel}, {van Beuzekom}, {van den
  Brand}, {van den Broeck}, {Vander-Hyde}, {van der Schaaf}, {van Heijningen},
  {Vanosky}, {van Putten}, {van Remortel}, {Vardaro}, {Vargas}, {Varma},
  {Vas{\'u}th}, {Vecchio}, {Vedovato}, {Veitch}, {Veitch}, {Venneberg},
  {Venugopalan}, {Verkindt}, {Verma}, {Verma}, {Veske}, {Vetrano},
  {Vicer{\'e}}, {Vidyant}, {Viets}, {Vijaykumar}, {Villa-Ortega}, {Vinet},
  {Virtuoso}, {Vitale}, {Vo}, {Vocca}, {von Reis}, {von Wrangel}, {Vorvick},
  {Vyatchanin}, {Wade}, {Wade}, {Wagner}, {Walet}, {Walker}, {Wallace},
  {Wallace}, {Walsh}, {Wang}, {Wang}, {Wang}, {Ward}, {Warner}, {Was},
  {Washimi}, {Washington}, {Watchi}, {Weaver}, {Webster}, {Weinert},
  {Weinstein}, {Weiss}, {Weller}, {Wellmann}, {Wen}, {We{\ss}els}, {Wette},
  {Whelan}, {White}, {Whiting}, {Whittle}, {Wilken}, {Williams}, {Williams},
  {Williamson}, {Willis}, {Willke}, {Wilson}, {Winkler}, {Wipf}, {Wlodarczyk},
  {Woan}, {Woehler}, {Wofford}, {Wong}, {Wu}, {Wu}, {Wu}, {Wu}, {Wysocki},
  {Xiao}, {Xu}, {Yamada}, {Yamamoto}, {Yamamoto}, {Yamamoto}, {Yamamoto},
  {Yamashita}, {Yamazaki}, {Yang}, {Yang}, {Yang}, {Yang}, {Yang}, {Yap},
  {Yeeles}, {Yelikar}, {Ying}, {Yokogawa}, {Yokoyama}, {Yokozawa}, {Yoo},
  {Yoshioka}, {Yu}, {Yu}, {Yuzurihara}, {Zadro{\.z}ny}, {Zanolin}, {Zeidler},
  {Zelenova}, {Zendri}, {Zevin}, {Zhan}, {Zhang}, {Zhang}, {Zhang}, {Zhang},
  {Zhang}, {Zhao}, {Zhao}, {Zhao}, {Zhao}, {Zheng}, {Zhou}, {Zhou}, {Zhu},
  {Zhu}, {Zimmerman}, {Zlochower}, {Zucker}, {Zweizig}, {LIGO Scientific
  Collaboration}, {VIRGO Collaboration}, \& {KAGRA
  Collaboration}}]{Abbott+2023}
{Abbott}, R., {Abbott}, T.~D., {Acernese}, F., {et~al.} 2023, Physical Review
  X, 13, 011048

\bibitem[{{Aerts} {et~al.}(2010){Aerts}, {Christensen-Dalsgaard}, \&
  {Kurtz}}]{Aerts+2010}
{Aerts}, C., {Christensen-Dalsgaard}, J., \& {Kurtz}, D.~W. 2010,
  {Asteroseismology} (Springerslowly pulsating B stars)

\bibitem[{{Aerts} {et~al.}(2019){Aerts}, {Mathis}, \& {Rogers}}]{Aerts+2019}
{Aerts}, C., {Mathis}, S., \& {Rogers}, T.~M. 2019, \araa, 57, 35

\bibitem[{{Aerts} \& {Tkachenko}(2023)}]{Aerts+2023:2023arXiv231108453A}
{Aerts}, C. \& {Tkachenko}, A. 2023, arXiv e-prints, arXiv:2311.08453

\bibitem[{{Angulo} {et~al.}(1999){Angulo}, {Arnould}, {Rayet}, {Descouvemont},
  {Baye}, {Leclercq-Willain}, {Coc}, {Barhoumi}, {Aguer}, {Rolfs}, {Kunz},
  {Hammer}, {Mayer}, {Paradellis}, {Kossionides}, {Chronidou}, {Spyrou},
  {degl'Innocenti}, {Fiorentini}, {Ricci}, {Zavatarelli}, {Providencia},
  {Wolters}, {Soares}, {Grama}, {Rahighi}, {Shotter}, \& {Lamehi
  Rachti}}]{Angulo1999}
{Angulo}, C., {Arnould}, M., {Rayet}, M., {et~al.} 1999, \nphysa, 656, 3

\bibitem[{{Astropy Collaboration} {et~al.}(2022){Astropy Collaboration},
  {Price-Whelan}, {Lim}, {Earl}, {Starkman}, {Bradley}, {Shupe}, {Patil},
  {Corrales}, {Brasseur}, {N{"o}the}, {Donath}, {Tollerud}, {Morris},
  {Ginsburg}, {Vaher}, {Weaver}, {Tocknell}, {Jamieson}, {van Kerkwijk},
  {Robitaille}, {Merry}, {Bachetti}, {G{"u}nther}, {Aldcroft},
  {Alvarado-Montes}, {Archibald}, {B{'o}di}, {Bapat}, {Barentsen}, {Baz{'a}n},
  {Biswas}, {Boquien}, {Burke}, {Cara}, {Cara}, {Conroy}, {Conseil}, {Craig},
  {Cross}, {Cruz}, {D'Eugenio}, {Dencheva}, {Devillepoix}, {Dietrich},
  {Eigenbrot}, {Erben}, {Ferreira}, {Foreman-Mackey}, {Fox}, {Freij}, {Garg},
  {Geda}, {Glattly}, {Gondhalekar}, {Gordon}, {Grant}, {Greenfield}, {Groener},
  {Guest}, {Gurovich}, {Handberg}, {Hart}, {Hatfield-Dodds}, {Homeier},
  {Hosseinzadeh}, {Jenness}, {Jones}, {Joseph}, {Kalmbach}, {Karamehmetoglu},
  {Ka{l}uszy{'n}ski}, {Kelley}, {Kern}, {Kerzendorf}, {Koch}, {Kulumani},
  {Lee}, {Ly}, {Ma}, {MacBride}, {Maljaars}, {Muna}, {Murphy}, {Norman},
  {O'Steen}, {Oman}, {Pacifici}, {Pascual}, {Pascual-Granado}, {Patil},
  {Perren}, {Pickering}, {Rastogi}, {Roulston}, {Ryan}, {Rykoff}, {Sabater},
  {Sakurikar}, {Salgado}, {Sanghi}, {Saunders}, {Savchenko}, {Schwardt},
  {Seifert-Eckert}, {Shih}, {Jain}, {Shukla}, {Sick}, {Simpson},
  {Singanamalla}, {Singer}, {Singhal}, {Sinha}, {Sip{H{o}}cz}, {Spitler},
  {Stansby}, {Streicher}, {{{S}}umak}, {Swinbank}, {Taranu}, {Tewary},
  {Tremblay}, {Val-Borro}, {Van Kooten}, {Vasovi{'c}}, {Verma}, {de Miranda
  Cardoso}, {Williams}, {Wilson}, {Winkel}, {Wood-Vasey}, {Xue}, {Yoachim},
  {Zhang}, {Zonca}, \& {Astropy Project Contributors}}]{astropy:2022}
{Astropy Collaboration}, {Price-Whelan}, A.~M., {Lim}, P.~L., {et~al.} 2022,
  apj, 935, 167

\bibitem[{{Astropy Collaboration} {et~al.}(2018){Astropy Collaboration},
  {Price-Whelan}, {Sip{\H{o}}cz}, {G{\"u}nther}, {Lim}, {Crawford}, {Conseil},
  {Shupe}, {Craig}, {Dencheva}, {Ginsburg}, {Vand erPlas}, {Bradley},
  {P{\'e}rez-Su{\'a}rez}, {de Val-Borro}, {Aldcroft}, {Cruz}, {Robitaille},
  {Tollerud}, {Ardelean}, {Babej}, {Bach}, {Bachetti}, {Bakanov}, {Bamford},
  {Barentsen}, {Barmby}, {Baumbach}, {Berry}, {Biscani}, {Boquien}, {Bostroem},
  {Bouma}, {Brammer}, {Bray}, {Breytenbach}, {Buddelmeijer}, {Burke},
  {Calderone}, {Cano Rodr{\'\i}guez}, {Cara}, {Cardoso}, {Cheedella}, {Copin},
  {Corrales}, {Crichton}, {D'Avella}, {Deil}, {Depagne}, {Dietrich}, {Donath},
  {Droettboom}, {Earl}, {Erben}, {Fabbro}, {Ferreira}, {Finethy}, {Fox},
  {Garrison}, {Gibbons}, {Goldstein}, {Gommers}, {Greco}, {Greenfield},
  {Groener}, {Grollier}, {Hagen}, {Hirst}, {Homeier}, {Horton}, {Hosseinzadeh},
  {Hu}, {Hunkeler}, {Ivezi{\'c}}, {Jain}, {Jenness}, {Kanarek}, {Kendrew},
  {Kern}, {Kerzendorf}, {Khvalko}, {King}, {Kirkby}, {Kulkarni}, {Kumar},
  {Lee}, {Lenz}, {Littlefair}, {Ma}, {Macleod}, {Mastropietro}, {McCully},
  {Montagnac}, {Morris}, {Mueller}, {Mumford}, {Muna}, {Murphy}, {Nelson},
  {Nguyen}, {Ninan}, {N{\"o}the}, {Ogaz}, {Oh}, {Parejko}, {Parley}, {Pascual},
  {Patil}, {Patil}, {Plunkett}, {Prochaska}, {Rastogi}, {Reddy Janga},
  {Sabater}, {Sakurikar}, {Seifert}, {Sherbert}, {Sherwood-Taylor}, {Shih},
  {Sick}, {Silbiger}, {Singanamalla}, {Singer}, {Sladen}, {Sooley},
  {Sornarajah}, {Streicher}, {Teuben}, {Thomas}, {Tremblay}, {Turner},
  {Terr{\'o}n}, {van Kerkwijk}, {de la Vega}, {Watkins}, {Weaver}, {Whitmore},
  {Woillez}, {Zabalza}, \& {Astropy Contributors}}]{astropy:2018}
{Astropy Collaboration}, {Price-Whelan}, A.~M., {Sip{\H{o}}cz}, B.~M., {et~al.}
  2018, \aj, 156, 123

\bibitem[{{Astropy Collaboration} {et~al.}(2013){Astropy Collaboration},
  {Robitaille}, {Tollerud}, {Greenfield}, {Droettboom}, {Bray}, {Aldcroft},
  {Davis}, {Ginsburg}, {Price-Whelan}, {Kerzendorf}, {Conley}, {Crighton},
  {Barbary}, {Muna}, {Ferguson}, {Grollier}, {Parikh}, {Nair}, {Unther},
  {Deil}, {Woillez}, {Conseil}, {Kramer}, {Turner}, {Singer}, {Fox}, {Weaver},
  {Zabalza}, {Edwards}, {Azalee Bostroem}, {Burke}, {Casey}, {Crawford},
  {Dencheva}, {Ely}, {Jenness}, {Labrie}, {Lim}, {Pierfederici}, {Pontzen},
  {Ptak}, {Refsdal}, {Servillat}, \& {Streicher}}]{astropy:2013}
{Astropy Collaboration}, {Robitaille}, T.~P., {Tollerud}, E.~J., {et~al.} 2013,
  \aap, 558, A33

\bibitem[{{Baade}(1982)}]{Baade1982}
{Baade}, D. 1982, \aap, 105, 65

\bibitem[{{Basu} {et~al.}(2012){Basu}, {Verner}, {Chaplin}, \&
  {Elsworth}}]{Basu+2012:2012ApJ...746...76B}
{Basu}, S., {Verner}, G.~A., {Chaplin}, W.~J., \& {Elsworth}, Y. 2012, \apj,
  746, 76

\bibitem[{{Belczynski} {et~al.}(2002){Belczynski}, {Kalogera}, \&
  {Bulik}}]{Belczynski+2002:2002ApJ...572..407B}
{Belczynski}, K., {Kalogera}, V., \& {Bulik}, T. 2002, \apj, 572, 407

\bibitem[{{Bellinger} {et~al.}(2023){Bellinger}, {de Mink}, {van Rossem}, \&
  {Justham}}]{Bellinger+2023:2023arXiv231100038B}
{Bellinger}, E.~P., {de Mink}, S.~E., {van Rossem}, W.~E., \& {Justham}, S.
  2023, arXiv e-prints, arXiv:2311.00038

\bibitem[{{Blouin} {et~al.}(2020){Blouin}, {Shaffer}, {Saumon}, \&
  {Starrett}}]{Blouin2020}
{Blouin}, S., {Shaffer}, N.~R., {Saumon}, D., \& {Starrett}, C.~E. 2020, \apj,
  899, 46

\bibitem[{{Bodensteiner} {et~al.}(2020){Bodensteiner}, {Shenar}, \&
  {Sana}}]{Bodensteiner+2020}
{Bodensteiner}, J., {Shenar}, T., \& {Sana}, H. 2020, \aap, 641, A42

\bibitem[{{Borucki} {et~al.}(2010){Borucki}, {Koch}, {Basri}, {Batalha},
  {Brown}, {Caldwell}, {Caldwell}, {Christensen-Dalsgaard}, {Cochran},
  {DeVore}, {Dunham}, {Dupree}, {Gautier}, {Geary}, {Gilliland}, {Gould},
  {Howell}, {Jenkins}, {Kondo}, {Latham}, {Marcy}, {Meibom}, {Kjeldsen},
  {Lissauer}, {Monet}, {Morrison}, {Sasselov}, {Tarter}, {Boss}, {Brownlee},
  {Owen}, {Buzasi}, {Charbonneau}, {Doyle}, {Fortney}, {Ford}, {Holman},
  {Seager}, {Steffen}, {Welsh}, {Rowe}, {Anderson}, {Buchhave}, {Ciardi},
  {Walkowicz}, {Sherry}, {Horch}, {Isaacson}, {Everett}, {Fischer}, {Torres},
  {Johnson}, {Endl}, {MacQueen}, {Bryson}, {Dotson}, {Haas}, {Kolodziejczak},
  {Van Cleve}, {Chandrasekaran}, {Twicken}, {Quintana}, {Clarke}, {Allen},
  {Li}, {Wu}, {Tenenbaum}, {Verner}, {Bruhweiler}, {Barnes}, \&
  {Prsa}}]{Borucki+2010}
{Borucki}, W.~J., {Koch}, D., {Basri}, G., {et~al.} 2010, Science, 327, 977

\bibitem[{{Bouabid} {et~al.}(2013){Bouabid}, {Dupret}, {Salmon},
  {Montalb{\'a}n}, {Miglio}, \& et~al.}]{Bouabid+2013}
{Bouabid}, M.~P., {Dupret}, M.~A., {Salmon}, S., {et~al.} 2013, \mnras, 429,
  2500

\bibitem[{{Braun} \& {Langer}(1995)}]{Braun+1995}
{Braun}, H. \& {Langer}, N. 1995, \aap, 297, 483

\bibitem[{{Broekgaarden} {et~al.}(2022){Broekgaarden}, {Berger}, {Stevenson},
  {Justham}, {Mandel}, \& et~al.}]{Broekgaarden+2022}
{Broekgaarden}, F.~S., {Berger}, E., {Stevenson}, S., {et~al.} 2022, \mnras,
  516, 5737

\bibitem[{{Brunt}(1927)}]{BVF-brunt}
{Brunt}, D. 1927, Quarterly Journal of the Royal Meteorological Society, 53, 30

\bibitem[{{Burssens} {et~al.}(2023){Burssens}, {Bowman}, {Michielsen},
  {Sim{\'o}n-D{\'\i}az}, {Aerts}, {Vanlaer}, {Banyard}, {Nardetto}, {Townsend},
  {Handler}, {Mombarg}, {Vanderspek}, \& {Ricker}}]{Burssens+2023}
{Burssens}, S., {Bowman}, D.~M., {Michielsen}, M., {et~al.} 2023, Nature
  Astronomy, 7, 913

\bibitem[{{Buysschaert} {et~al.}(2018){Buysschaert}, {Aerts}, {Bowman},
  {Johnston}, {Van Reeth}, {Pedersen}, {Mathis}, \&
  {Neiner}}]{Buysschaert+2018:2018A&A...616A.148B}
{Buysschaert}, B., {Aerts}, C., {Bowman}, D.~M., {et~al.} 2018, \aap, 616, A148

\bibitem[{{Cantiello} {et~al.}(2007){Cantiello}, {Yoon}, {Langer}, \&
  {Livio}}]{Cantiello+2007}
{Cantiello}, M., {Yoon}, S.~C., {Langer}, N., \& {Livio}, M. 2007, \aap, 465,
  L29

\bibitem[{{Cassisi} {et~al.}(2007){Cassisi}, {Potekhin}, {Pietrinferni},
  {Catelan}, \& {Salaris}}]{Cassisi2007}
{Cassisi}, S., {Potekhin}, A.~Y., {Pietrinferni}, A., {Catelan}, M., \&
  {Salaris}, M. 2007, \apj, 661, 1094

\bibitem[{{Chen} {et~al.}(2021){Chen}, {Zhang}, {Li}, {Luo}, {Li}, {Su}, \&
  {Chen}}]{Chen2021}
{Chen}, X., {Zhang}, X., {Li}, Y., {et~al.} 2021, \apj, 920, 76

\bibitem[{{Christensen-Dalsgaard}(2008)}]{Christensen-Dalsgaard+2008}
{Christensen-Dalsgaard}, J. 2008, \apss, 316, 113

\bibitem[{{Chugunov} {et~al.}(2007){Chugunov}, {Dewitt}, \&
  {Yakovlev}}]{Chugunov2007}
{Chugunov}, A.~I., {Dewitt}, H.~E., \& {Yakovlev}, D.~G. 2007, \prd, 76, 025028

\bibitem[{{Claeys} {et~al.}(2011){Claeys}, {de Mink}, {Pols}, {Eldridge}, \&
  {Baes}}]{Claeys+2011:2011A&A...528A.131C}
{Claeys}, J.~S.~W., {de Mink}, S.~E., {Pols}, O.~R., {Eldridge}, J.~J., \&
  {Baes}, M. 2011, \aap, 528, A131

\bibitem[{{Claret} \& {Torres}(2017)}]{Claret+2017}
{Claret}, A. \& {Torres}, G. 2017, \apj, 849, 18

\bibitem[{{Cox} {et~al.}(1992){Cox}, {Morgan}, {Rogers}, \&
  {Iglesias}}]{Cox+1992}
{Cox}, A.~N., {Morgan}, S.~M., {Rogers}, F.~J., \& {Iglesias}, C.~A. 1992,
  \apj, 393, 272

\bibitem[{{Crowe} \& {Matalas}(1982)}]{Crowe+1982}
{Crowe}, R.~A. \& {Matalas}, R. 1982, \aap, 108, 55

\bibitem[{{Crowther}(2007)}]{Crowther2007}
{Crowther}, P.~A. 2007, \araa, 45, 177

\bibitem[{{Cyburt} {et~al.}(2010){Cyburt}, {Amthor}, {Ferguson}, {Meisel},
  {Smith}, {Warren}, {Heger}, {Hoffman}, {Rauscher}, {Sakharuk}, {Schatz},
  {Thielemann}, \& {Wiescher}}]{Cyburt2010}
{Cyburt}, R.~H., {Amthor}, A.~M., {Ferguson}, R., {et~al.} 2010, \apjs, 189,
  240

\bibitem[{{de Mink} {et~al.}(2013){de Mink}, {Langer}, {Izzard}, {Sana}, \& {de
  Koter}}]{deMink+2013:2013ApJ...764..166D}
{de Mink}, S.~E., {Langer}, N., {Izzard}, R.~G., {Sana}, H., \& {de Koter}, A.
  2013, \apj, 764, 166

\bibitem[{{de Mink} {et~al.}(2007){de Mink}, {Pols}, \&
  {Hilditch}}]{deMink+2007:2007A&A...467.1181D}
{de Mink}, S.~E., {Pols}, O.~R., \& {Hilditch}, R.~W. 2007, \aap, 467, 1181

\bibitem[{{de Mink} {et~al.}(2014){de Mink}, {Sana}, {Langer}, {Izzard}, \&
  {Schneider}}]{deMink+2014}
{de Mink}, S.~E., {Sana}, H., {Langer}, N., {Izzard}, R.~G., \& {Schneider},
  F.~R.~N. 2014, \apj, 782, 7

\bibitem[{{Dervi{\c s}o{\u g}lu} {et~al.}(2010){Dervi{\c s}o{\u g}lu}, {Tout},
  \& {Ibano{\u g}lu}}]{Dervisoglu+2010}
{Dervi{\c s}o{\u g}lu}, A., {Tout}, C.~A., \& {Ibano{\u g}lu}, C. 2010, \mnras,
  406, 1071

\bibitem[{{Dominik} {et~al.}(2015){Dominik}, {Berti}, {O'Shaughnessy},
  {Mandel}, {Belczynski}, {Fryer}, {Holz}, {Bulik}, \&
  {Pannarale}}]{Dominik+2015:2015ApJ...806..263D}
{Dominik}, M., {Berti}, E., {O'Shaughnessy}, R., {et~al.} 2015, \apj, 806, 263

\bibitem[{{Duch{\^e}ne} \& {Kraus}(2013)}]{Duchene+2013:2013ARA&A..51..269D}
{Duch{\^e}ne}, G. \& {Kraus}, A. 2013, \araa, 51, 269

\bibitem[{{Dziembowski}(1971)}]{Dziembowski+1971}
{Dziembowski}, W.~A. 1971, \actaa, 21, 289

\bibitem[{{Dziembowski} {et~al.}(1993){Dziembowski}, {Moskalik}, \&
  {Pamyatnykh}}]{Dziembowski1993}
{Dziembowski}, W.~A., {Moskalik}, P., \& {Pamyatnykh}, A.~A. 1993, \mnras, 265,
  588

\bibitem[{{Eggleton}(1983)}]{Eggleton1983}
{Eggleton}, P.~P. 1983, \apj, 268, 368

\bibitem[{{Eldridge} {et~al.}(2013){Eldridge}, {Fraser}, {Smartt}, {Maund}, \&
  {Crockett}}]{Eldridge+2013:2013MNRAS.436..774E}
{Eldridge}, J.~J., {Fraser}, M., {Smartt}, S.~J., {Maund}, J.~R., \&
  {Crockett}, R.~M. 2013, \mnras, 436, 774

\bibitem[{{Ferguson} {et~al.}(2005){Ferguson}, {Alexander}, {Allard}, {Barman},
  {Bodnarik}, {Hauschildt}, {Heffner-Wong}, \& {Tamanai}}]{Ferguson2005}
{Ferguson}, J.~W., {Alexander}, D.~R., {Allard}, F., {et~al.} 2005, \apj, 623,
  585

\bibitem[{{Fragos} {et~al.}(2013){Fragos}, {Lehmer}, {Tremmel}, {Tzanavaris},
  {Basu-Zych}, {Belczynski}, {Hornschemeier}, {Jenkins}, {Kalogera}, {Ptak}, \&
  {Zezas}}]{Fragos+2013:2013ApJ...764...41F}
{Fragos}, T., {Lehmer}, B., {Tremmel}, M., {et~al.} 2013, \apj, 764, 41

\bibitem[{{Fuller} {et~al.}(1985){Fuller}, {Fowler}, \& {Newman}}]{Fuller1985}
{Fuller}, G.~M., {Fowler}, W.~A., \& {Newman}, M.~J. 1985, \apj, 293, 1

\bibitem[{{Goldstein} \& {Townsend}(2020)}]{Goldstein+2020}
{Goldstein}, J. \& {Townsend}, R.~H.~D. 2020, \apj, 899, 116

\bibitem[{{G{\"o}tberg} {et~al.}(2017){G{\"o}tberg}, {de Mink}, \&
  {Groh}}]{Gotberg+2017:2017A&A...608A..11G}
{G{\"o}tberg}, Y., {de Mink}, S.~E., \& {Groh}, J.~H. 2017, \aap, 608, A11

\bibitem[{{G{\"o}tberg} {et~al.}(2020){G{\"o}tberg}, {Korol}, {Lamberts},
  {Kupfer}, {Breivik}, {Ludwig}, \& {Drout}}]{Gotberg+2020}
{G{\"o}tberg}, Y., {Korol}, V., {Lamberts}, A., {et~al.} 2020, \apj, 904, 56

\bibitem[{{Guo} {et~al.}(2017{\natexlab{a}}){Guo}, {Gies}, \&
  {Matson}}]{Guo2017b}
{Guo}, Z., {Gies}, D.~R., \& {Matson}, R.~A. 2017{\natexlab{a}}, \apj, 851, 39

\bibitem[{{Guo} {et~al.}(2017{\natexlab{b}}){Guo}, {Gies}, {Matson},
  {Garc{\'\i}a Hern{\'a}ndez}, {Han}, \& {Chen}}]{Guo+2017:2017ApJ...837..114G}
{Guo}, Z., {Gies}, D.~R., {Matson}, R.~A., {et~al.} 2017{\natexlab{b}}, \apj,
  837, 114

\bibitem[{{Guo} \& {Li}(2019)}]{Guo2019}
{Guo}, Z. \& {Li}, G. 2019, \apjl, 882, L5

\bibitem[{{Guzik} {et~al.}(2000){Guzik}, {Kaye}, {Bradley}, {Cox}, \&
  {Neuforge}}]{Guzik+2000}
{Guzik}, J.~A., {Kaye}, A.~B., {Bradley}, P.~A., {Cox}, A.~N., \& {Neuforge},
  C. 2000, \apjl, 542, L57

\bibitem[{Harris {et~al.}(2020)Harris, Millman, van~der Walt, Gommers,
  Virtanen, Cournapeau, Wieser, Taylor, Berg, Smith, Kern, Picus, Hoyer, van
  Kerkwijk, Brett, Haldane, del R{\'{i}}o, Wiebe, Peterson,
  G{\'{e}}rard-Marchant, Sheppard, Reddy, Weckesser, Abbasi, Gohlke, \&
  Oliphant}]{numpy}
Harris, C.~R., Millman, K.~J., van~der Walt, S.~J., {et~al.} 2020, Nature, 585,
  357

\bibitem[{{Hatta}(2023)}]{Hatta+2023}
{Hatta}, Y. 2023, \apj, 950, 165

\bibitem[{{Heber}(2016)}]{Heber2016}
{Heber}, U. 2016, \pasp, 128, 082001

\bibitem[{{Hellings}(1983)}]{Hellings1983}
{Hellings}, P. 1983, \apss, 96, 37

\bibitem[{{Herwig}(2000)}]{Herwig+2000}
{Herwig}, F. 2000, \aap, 360, 952

\bibitem[{Hunter(2007)}]{matplotlib}
Hunter, J.~D. 2007, Computing in Science \& Engineering, 9, 90

\bibitem[{{Iglesias} \& {Rogers}(1993)}]{Iglesias1993}
{Iglesias}, C.~A. \& {Rogers}, F.~J. 1993, \apj, 412, 752

\bibitem[{{Iglesias} \& {Rogers}(1996)}]{Iglesias1996}
{Iglesias}, C.~A. \& {Rogers}, F.~J. 1996, \apj, 464, 943

\bibitem[{{Iorio} {et~al.}(2023){Iorio}, {Mapelli}, {Costa}, {Spera},
  {Escobar}, {Sgalletta}, {Trani}, {Korb}, {Santoliquido}, {Dall'Amico},
  {Gaspari}, \& {Bressan}}]{Iorio+2023}
{Iorio}, G., {Mapelli}, M., {Costa}, G., {et~al.} 2023, \mnras, 524, 426

\bibitem[{{Irwin}(2004)}]{Irwin2004}
{Irwin}, A.~W. 2004, The FreeEOS Code for Calculating the Equation of State for
  Stellar Interiors

\bibitem[{{Itoh} {et~al.}(1996){Itoh}, {Hayashi}, {Nishikawa}, \&
  {Kohyama}}]{Itoh1996}
{Itoh}, N., {Hayashi}, H., {Nishikawa}, A., \& {Kohyama}, Y. 1996, \apjs, 102,
  411

\bibitem[{{Jermyn} {et~al.}(2023){Jermyn}, {Bauer}, {Schwab}, {Farmer}, {Ball},
  {Bellinger}, {Dotter}, {Joyce}, {Marchant}, {Mombarg}, {Wolf}, {Sunny Wong},
  {Cinquegrana}, {Farrell}, {Smolec}, {Thoul}, {Cantiello}, {Herwig}, {Toloza},
  {Bildsten}, {Townsend}, \& {Timmes}}]{Jermyn2023}
{Jermyn}, A.~S., {Bauer}, E.~B., {Schwab}, J., {et~al.} 2023, \apjs, 265, 15

\bibitem[{{Jermyn} {et~al.}(2021){Jermyn}, {Schwab}, {Bauer}, {Timmes}, \&
  {Potekhin}}]{Jermyn2021}
{Jermyn}, A.~S., {Schwab}, J., {Bauer}, E., {Timmes}, F.~X., \& {Potekhin},
  A.~Y. 2021, \apj, 913, 72

\bibitem[{{Johnston}(2021)}]{Johnston+2021}
{Johnston}, C. 2021, \aap, 655, A29

\bibitem[{{Johnston} {et~al.}(2019{\natexlab{a}}){Johnston}, {Pavlovski}, \&
  {Tkachenko}}]{Johnston2019a}
{Johnston}, C., {Pavlovski}, K., \& {Tkachenko}, A. 2019{\natexlab{a}}, \aap,
  628, A25

\bibitem[{{Johnston} {et~al.}(2019{\natexlab{b}}){Johnston}, {Tkachenko},
  {Aerts}, {Molenberghs}, {Bowman}, {Pedersen}, {Buysschaert}, \&
  {P{\'a}pics}}]{Johnston2019b}
{Johnston}, C., {Tkachenko}, A., {Aerts}, C., {et~al.} 2019{\natexlab{b}},
  \mnras, 482, 1231

\bibitem[{{Kippenhahn}(1969)}]{Kippenhahn+1969:1969A&A.....3...83K}
{Kippenhahn}, R. 1969, \aap, 3, 83

\bibitem[{{Kippenhahn} {et~al.}(1980){Kippenhahn}, {Ruschenplatt}, \&
  {Thomas}}]{Kippenhahn+1980}
{Kippenhahn}, R., {Ruschenplatt}, G., \& {Thomas}, H.~C. 1980, \aap, 91, 175

\bibitem[{{Kolb} \& {Ritter}(1990)}]{Kolb+1990}
{Kolb}, U. \& {Ritter}, H. 1990, \aap, 236, 385

\bibitem[{{Kurtz}(2022)}]{Kurtz+2022}
{Kurtz}, D.~W. 2022, \araa, 60, 31

\bibitem[{{Labadie-Bartz} {et~al.}(2022){Labadie-Bartz}, {Carciofi}, {Henrique
  de Amorim}, {Rubio}, {Luiz Figueiredo}, {Ticiani dos Santos}, \&
  {Thomson-Paressant}}]{LabadieBartz2022}
{Labadie-Bartz}, J., {Carciofi}, A.~C., {Henrique de Amorim}, T., {et~al.}
  2022, \aj, 163, 226

\bibitem[{{Langanke} \& {Mart{\'{\i}}nez-Pinedo}(2000)}]{Langanke2000}
{Langanke}, K. \& {Mart{\'{\i}}nez-Pinedo}, G. 2000, Nuclear Physics A, 673,
  481

\bibitem[{{Langer} {et~al.}(1983){Langer}, {Fricke}, \&
  {Sugimoto}}]{Langer+1983}
{Langer}, N., {Fricke}, K.~J., \& {Sugimoto}, D. 1983, \aap, 126, 207

\bibitem[{{Lau} {et~al.}(2014){Lau}, {Izzard}, \&
  {Schneider}}]{Lau+2014:2014A&A...570A.125L}
{Lau}, H.~H.~B., {Izzard}, R.~G., \& {Schneider}, F.~R.~N. 2014, \aap, 570,
  A125

\bibitem[{{Lau} {et~al.}(2024){Lau}, {Hirai}, {Mandel}, \&
  {Tout}}]{Lau+2024:2024arXiv240109570L}
{Lau}, M. Y.~M., {Hirai}, R., {Mandel}, I., \& {Tout}, C.~A. 2024, arXiv
  e-prints, arXiv:2401.09570

\bibitem[{{Ledoux}(1947)}]{Ledoux+1947}
{Ledoux}, P. 1947, \apj, 105, 305

\bibitem[{{Maeder} \& {Meynet}(2000)}]{Maeder2000}
{Maeder}, A. \& {Meynet}, G. 2000, \araa, 38, 143

\bibitem[{{Marchant} {et~al.}(2021){Marchant}, {Pappas}, {Gallegos-Garcia},
  {Berry}, {Taam}, {Kalogera}, \& {Podsiadlowski}}]{Marchant2021+}
{Marchant}, P., {Pappas}, K. M.~W., {Gallegos-Garcia}, M., {et~al.} 2021, \aap,
  650, A107

\bibitem[{{McClelland} \&
  {Eldridge}(2016)}]{McClelland+2016:2016MNRAS.459.1505M}
{McClelland}, L.~A.~S. \& {Eldridge}, J.~J. 2016, \mnras, 459, 1505

\bibitem[{{Michielsen} {et~al.}(2021){Michielsen}, {Aerts}, \&
  {Bowman}}]{Michielsen+2021}
{Michielsen}, M., {Aerts}, C., \& {Bowman}, D.~M. 2021, \aap, 650, A175

\bibitem[{{Miglio} {et~al.}(2008){Miglio}, {Montalb{\'a}n}, {Eggenberger}, \&
  {Noels}}]{Miglio+2008}
{Miglio}, A., {Montalb{\'a}n}, J., {Eggenberger}, P., \& {Noels}, A. 2008,
  Astronomische Nachrichten, 329, 529

\bibitem[{{Miszuda} {et~al.}(2022){Miszuda}, {Ko{\l}aczek-Szyma{\'n}ski},
  {Szewczuk}, \&
  {Daszy{\'n}ska-Daszkiewicz}}]{Miszuda+2022:2022MNRAS.514..622M}
{Miszuda}, A., {Ko{\l}aczek-Szyma{\'n}ski}, P.~A., {Szewczuk}, W., \&
  {Daszy{\'n}ska-Daszkiewicz}, J. 2022, \mnras, 514, 622

\bibitem[{{Miszuda} {et~al.}(2021){Miszuda}, {Szewczuk}, \&
  {Daszy{\'n}ska-Daszkiewicz}}]{Miszuda+2021}
{Miszuda}, A., {Szewczuk}, W., \& {Daszy{\'n}ska-Daszkiewicz}, J. 2021, \mnras,
  505, 3206

\bibitem[{{Mitalas}(1972)}]{Mitalas+1972}
{Mitalas}, R. 1972, \apj, 177, 693

\bibitem[{{Moe} \& {Di Stefano}(2017)}]{Moe+2017}
{Moe}, M. \& {Di Stefano}, R. 2017, \apjs, 230, 15

\bibitem[{{Mombarg}(2023)}]{Mombarg2023}
{Mombarg}, J.~S.~G. 2023, \aap, 677, A63

\bibitem[{{Mombarg} {et~al.}(2019){Mombarg}, {Van Reeth}, {Pedersen},
  {Molenberghs}, {Bowman}, {Johnston}, {Tkachenko}, \& {Aerts}}]{Mombarg2019}
{Mombarg}, J.~S.~G., {Van Reeth}, T., {Pedersen}, M.~G., {et~al.} 2019, \mnras,
  485, 3248

\bibitem[{{Morton}(1960)}]{Morton+1960:1960ApJ...132..146M}
{Morton}, D.~C. 1960, \apj, 132, 146

\bibitem[{{Moyano} {et~al.}(2024){Moyano}, {Eggenberger}, \&
  {Salmon}}]{Moyano2024}
{Moyano}, F.~D., {Eggenberger}, P., \& {Salmon}, S.~J.~A.~J. 2024, \aap, 681,
  L16

\bibitem[{{Neo} {et~al.}(1977){Neo}, {Miyaji}, {Nomoto}, \&
  {Sugimoto}}]{Neo+1977}
{Neo}, S., {Miyaji}, S., {Nomoto}, K., \& {Sugimoto}, D. 1977, \pasj, 29, 249

\bibitem[{{Oda} {et~al.}(1994){Oda}, {Hino}, {Muto}, {Takahara}, \&
  {Sato}}]{Oda1994}
{Oda}, T., {Hino}, M., {Muto}, K., {Takahara}, M., \& {Sato}, K. 1994, Atomic
  Data and Nuclear Data Tables, 56, 231

\bibitem[{{Offner} {et~al.}(2023){Offner}, {Moe}, {Kratter}, {Sadavoy},
  {Jensen}, \& {Tobin}}]{Offner+2023:2023ASPC..534..275O}
{Offner}, S.~S.~R., {Moe}, M., {Kratter}, K.~M., {et~al.} 2023, in Astronomical
  Society of the Pacific Conference Series, Vol. 534, Protostars and Planets
  VII, ed. S.~{Inutsuka}, Y.~{Aikawa}, T.~{Muto}, K.~{Tomida}, \& M.~{Tamura},
  275

\bibitem[{{Ouazzani} {et~al.}(2020){Ouazzani}, {Ligni{\`e}res}, {Dupret},
  {Salmon}, {Ballot}, {Christophe}, \& {Takata}}]{Ouazzani2020}
{Ouazzani}, R.~M., {Ligni{\`e}res}, F., {Dupret}, M.~A., {et~al.} 2020, \aap,
  640, A49

\bibitem[{{Packet}(1981)}]{Packet+1981}
{Packet}, W. 1981, \aap, 102, 17

\bibitem[{{Paczy{\'n}ski}(1966)}]{Paczynski+1966:1966AcA....16..231P}
{Paczy{\'n}ski}, B. 1966, \actaa, 16, 231

\bibitem[{{Paczynski}(1991)}]{Paczynski+1991}
{Paczynski}, B. 1991, \apj, 370, 597

\bibitem[{{Pamyatnykh}(1999)}]{Pamyatnykh+1999}
{Pamyatnykh}, A.~A. 1999, \actaa, 49, 119

\bibitem[{pandas~development team(2022)}]{pandas_1.4.2}
pandas~development team, T. 2022, pandas-dev/pandas: Pandas 1.4.2

\bibitem[{Paxton(2023)}]{mesa_zenodo}
Paxton, B. 2023, {Modules for Experiments in Stellar Astrophysics (MESA)}

\bibitem[{{Paxton} {et~al.}(2011){Paxton}, {Bildsten}, {Dotter}, {Herwig},
  {Lesaffre}, \& {Timmes}}]{Paxton2011}
{Paxton}, B., {Bildsten}, L., {Dotter}, A., {et~al.} 2011, \apjs, 192, 3

\bibitem[{{Paxton} {et~al.}(2013){Paxton}, {Cantiello}, {Arras}, {Bildsten},
  {Brown}, {Dotter}, {Mankovich}, {Montgomery}, {Stello}, {Timmes}, \&
  {Townsend}}]{Paxton2013}
{Paxton}, B., {Cantiello}, M., {Arras}, P., {et~al.} 2013, \apjs, 208, 4

\bibitem[{{Paxton} {et~al.}(2015){Paxton}, {Marchant}, {Schwab}, {Bauer},
  {Bildsten}, {Cantiello}, {Dessart}, {Farmer}, {Hu}, {Langer}, {Townsend},
  {Townsley}, \& {Timmes}}]{Paxton2015}
{Paxton}, B., {Marchant}, P., {Schwab}, J., {et~al.} 2015, \apjs, 220, 15

\bibitem[{{Paxton} {et~al.}(2018){Paxton}, {Schwab}, {Bauer}, {Bildsten},
  {Blinnikov}, {Duffell}, {Farmer}, {Goldberg}, {Marchant}, {Sorokina},
  {Thoul}, {Townsend}, \& {Timmes}}]{Paxton2018}
{Paxton}, B., {Schwab}, J., {Bauer}, E.~B., {et~al.} 2018, \apjs, 234, 34

\bibitem[{{Paxton} {et~al.}(2019){Paxton}, {Smolec}, {Schwab}, {Gautschy},
  {Bildsten}, {Cantiello}, {Dotter}, {Farmer}, {Goldberg}, {Jermyn}, {Kanbur},
  {Marchant}, {Thoul}, {Townsend}, {Wolf}, {Zhang}, \& {Timmes}}]{Paxton2019}
{Paxton}, B., {Smolec}, R., {Schwab}, J., {et~al.} 2019, \apjs, 243, 10

\bibitem[{{Pedersen}(2022)}]{Pedersen+2022}
{Pedersen}, M.~G. 2022, \apj, 930, 94

\bibitem[{{Pedersen} {et~al.}(2018){Pedersen}, {Aerts}, {P{\'a}pics}, \&
  {Rogers}}]{Pedersen+2018}
{Pedersen}, M.~G., {Aerts}, C., {P{\'a}pics}, P.~I., \& {Rogers}, T.~M. 2018,
  \aap, 614, A128

\bibitem[{{Petrovic} {et~al.}(2005){Petrovic}, {Langer}, \& {van der
  Hucht}}]{Petrovic+2005}
{Petrovic}, J., {Langer}, N., \& {van der Hucht}, K.~A. 2005, \aap, 435, 1013

\bibitem[{{Podsiadlowski} {et~al.}(1992){Podsiadlowski}, {Joss}, \&
  {Hsu}}]{Podsiadlowski+1992:1992ApJ...391..246P}
{Podsiadlowski}, P., {Joss}, P.~C., \& {Hsu}, J.~J.~L. 1992, \apj, 391, 246

\bibitem[{{Popham} \& {Narayan}(1991)}]{Popham+1991}
{Popham}, R. \& {Narayan}, R. 1991, \apj, 370, 604

\bibitem[{{Potekhin} \& {Chabrier}(2010)}]{Potekhin2010}
{Potekhin}, A.~Y. \& {Chabrier}, G. 2010, Contributions to Plasma Physics, 50,
  82

\bibitem[{{Poutanen}(2017)}]{Poutanen2017}
{Poutanen}, J. 2017, \apj, 835, 119

\bibitem[{{Renzo} \& {G{\"o}tberg}(2021)}]{Renzo+2021}
{Renzo}, M. \& {G{\"o}tberg}, Y. 2021, \apj, 923, 277

\bibitem[{{Renzo} {et~al.}(2019){Renzo}, {Zapartas}, {de Mink}, {G{\"o}tberg},
  {Justham}, {Farmer}, {Izzard}, {Toonen}, \&
  {Sana}}]{Renzo+2019:2019A&A...624A..66R}
{Renzo}, M., {Zapartas}, E., {de Mink}, S.~E., {et~al.} 2019, \aap, 624, A66

\bibitem[{{Renzo} {et~al.}(2023){Renzo}, {Zapartas}, {Justham}, {Breivik},
  {Lau}, {Farmer}, {Cantiello}, \& {Metzger}}]{Renzo+2023}
{Renzo}, M., {Zapartas}, E., {Justham}, S., {et~al.} 2023, \apjl, 942, L32

\bibitem[{{Ricker} {et~al.}(2015){Ricker}, {Winn}, {Vanderspek}, {Latham},
  {Bakos}, {Bean}, {Berta-Thompson}, {Brown}, {Buchhave}, {Butler}, {Butler},
  {Chaplin}, {Charbonneau}, {Christensen-Dalsgaard}, {Clampin}, {Deming},
  {Doty}, {De Lee}, {Dressing}, {Dunham}, {Endl}, {Fressin}, {Ge}, {Henning},
  {Holman}, {Howard}, {Ida}, {Jenkins}, {Jernigan}, {Johnson}, {Kaltenegger},
  {Kawai}, {Kjeldsen}, {Laughlin}, {Levine}, {Lin}, {Lissauer}, {MacQueen},
  {Marcy}, {McCullough}, {Morton}, {Narita}, {Paegert}, {Palle}, {Pepe},
  {Pepper}, {Quirrenbach}, {Rinehart}, {Sasselov}, {Sato}, {Seager},
  {Sozzetti}, {Stassun}, {Sullivan}, {Szentgyorgyi}, {Torres}, {Udry}, \&
  {Villasenor}}]{Ricker+2015}
{Ricker}, G.~R., {Winn}, J.~N., {Vanderspek}, R., {et~al.} 2015, Journal of
  Astronomical Telescopes, Instruments, and Systems, 1, 014003

\bibitem[{{Rogers} \& {Nayfonov}(2002)}]{Rogers2002}
{Rogers}, F.~J. \& {Nayfonov}, A. 2002, \apj, 576, 1064

\bibitem[{{Rui} \& {Fuller}(2021)}]{Rui+2021}
{Rui}, N.~Z. \& {Fuller}, J. 2021, \mnras, 508, 1618

\bibitem[{{Salmon} {et~al.}(2022){Salmon}, {Moyano}, {Eggenberger},
  {Haemmerl{\'e}}, \& {Buldgen}}]{Salmon2022}
{Salmon}, S.~J.~A.~J., {Moyano}, F.~D., {Eggenberger}, P., {Haemmerl{\'e}}, L.,
  \& {Buldgen}, G. 2022, \aap, 664, L1

\bibitem[{{Sana} {et~al.}(2012){Sana}, {de Mink}, {de Koter}, {Langer},
  {Evans}, \& et~al.}]{Sana+2012}
{Sana}, H., {de Mink}, S.~E., {de Koter}, A., {et~al.} 2012, Science, 337, 444

\bibitem[{{Saumon} {et~al.}(1995){Saumon}, {Chabrier}, \& {van
  Horn}}]{Saumon1995}
{Saumon}, D., {Chabrier}, G., \& {van Horn}, H.~M. 1995, \apjs, 99, 713

\bibitem[{{Sekaran} {et~al.}(2021){Sekaran}, {Tkachenko}, {Johnston}, \&
  {Aerts}}]{Sekaran2021}
{Sekaran}, S., {Tkachenko}, A., {Johnston}, C., \& {Aerts}, C. 2021, \aap, 648,
  A91

\bibitem[{{Shi} {et~al.}(2022){Shi}, {Qian}, \& {Li}}]{Shi2022}
{Shi}, X.-d., {Qian}, S.-b., \& {Li}, L.-J. 2022, \apjs, 259, 50

\bibitem[{{Silva Aguirre} {et~al.}(2011){Silva Aguirre}, {Ballot}, {Serenelli},
  \& {Weiss}}]{SilvaAguirre+2011}
{Silva Aguirre}, V., {Ballot}, J., {Serenelli}, A.~M., \& {Weiss}, A. 2011,
  \aap, 529, A63

\bibitem[{{Smak}(1962)}]{Smak+1962:1962AcA....12...28S}
{Smak}, J. 1962, \actaa, 12, 28

\bibitem[{{Staritsin}(2019)}]{Staritsin+2019}
{Staritsin}, E.~I. 2019, \apss, 364, 110

\bibitem[{{Sun} {et~al.}(2023){Sun}, {Townsend}, \& {Guo}}]{Sun+2023}
{Sun}, M., {Townsend}, R.~H.~D., \& {Guo}, Z. 2023, \apj, 945, 43

\bibitem[{{Tassoul}(1980)}]{Tassoul+1980}
{Tassoul}, M. 1980, \apjs, 43, 469

\bibitem[{{Tauris} {et~al.}(2015){Tauris}, {Langer}, \&
  {Podsiadlowski}}]{Tauris+2015:2015MNRAS.451.2123T}
{Tauris}, T.~M., {Langer}, N., \& {Podsiadlowski}, P. 2015, \mnras, 451, 2123

\bibitem[{{Timmes} \& {Swesty}(2000)}]{Timmes2000}
{Timmes}, F.~X. \& {Swesty}, F.~D. 2000, \apjs, 126, 501

\bibitem[{{Toonen} \& {Nelemans}(2013)}]{Toonen+2013}
{Toonen}, S. \& {Nelemans}, G. 2013, \aap, 557, A87

\bibitem[{{Torres} {et~al.}(2010){Torres}, {Andersen}, \&
  {Gim{\'e}nez}}]{Torres2010}
{Torres}, G., {Andersen}, J., \& {Gim{\'e}nez}, A. 2010, \aapr, 18, 67

\bibitem[{{Townsend}(2003)}]{Townsend2003}
{Townsend}, R.~H.~D. 2003, \mnras, 343, 125

\bibitem[{{Townsend} {et~al.}(2018){Townsend}, {Goldstein}, \&
  {Zweibel}}]{Townsend+2018}
{Townsend}, R.~H.~D., {Goldstein}, J., \& {Zweibel}, E.~G. 2018, \mnras, 475,
  879

\bibitem[{{Townsend} \& {Teitler}(2013)}]{Townsend+2013}
{Townsend}, R.~H.~D. \& {Teitler}, S.~A. 2013, \mnras, 435, 3406

\bibitem[{{V{\"a}is{\"a}l{\"a}}(1925)}]{BVF-vaisala}
{V{\"a}is{\"a}l{\"a}}, V. 1925, Societus Scientiarum Fennica Commentationes
  Physico-Mathematicae, 2, 19

\bibitem[{{van der Linden}(1987)}]{vanderLinden+1987:1987A&A...178..170V}
{van der Linden}, T.~J. 1987, \aap, 178, 170

\bibitem[{Van~Rossum \& Drake(2009)}]{python}
Van~Rossum, G. \& Drake, F.~L. 2009, Python 3 Reference Manual (Scotts Valley,
  CA: CreateSpace)

\bibitem[{{van Son} {et~al.}(2022){van Son}, {de Mink}, {Renzo}, {Justham},
  {Zapartas}, {Breivik}, {Callister}, {Farr}, \&
  {Conroy}}]{vanSon+2022:2022ApJ...940..184V}
{van Son}, L.~A.~C., {de Mink}, S.~E., {Renzo}, M., {et~al.} 2022, \apj, 940,
  184

\bibitem[{{Virtanen} {et~al.}(2020){Virtanen}, {Gommers}, {Oliphant},
  {Haberland}, {Reddy}, \& et~al.}]{Virtanen+2020}
{Virtanen}, P., {Gommers}, R., {Oliphant}, T.~E., {et~al.} 2020, Nature
  Methods, 17, 261

\bibitem[{{Waelkens}(1991)}]{Waelkens+1991}
{Waelkens}, C. 1991, \aap, 246, 453

\bibitem[{{Waelkens} \& {Rufener}(1985)}]{Waelkens+1985}
{Waelkens}, C. \& {Rufener}, F. 1985, \aap, 152, 6

\bibitem[{{W}es {M}c{K}inney(2010)}]{pandas_paper}
{W}es {M}c{K}inney. 2010, in {P}roceedings of the 9th {P}ython in {S}cience
  {C}onference, ed. {S}t\'efan van~der {W}alt \& {J}arrod {M}illman, 56 -- 61

\bibitem[{{Xin} {et~al.}(2022){Xin}, {Renzo}, \& {Metzger}}]{Xin+2022}
{Xin}, C., {Renzo}, M., \& {Metzger}, B.~D. 2022, \mnras, 516, 5816

\bibitem[{{Yoon} {et~al.}(2017){Yoon}, {Dessart}, \&
  {Clocchiatti}}]{Yoon+2017:2017ApJ...840...10Y}
{Yoon}, S.-C., {Dessart}, L., \& {Clocchiatti}, A. 2017, \apj, 840, 10

\bibitem[{{Yoon} {et~al.}(2010){Yoon}, {Woosley}, \&
  {Langer}}]{Yoon+2010:2010ApJ...725..940Y}
{Yoon}, S.~C., {Woosley}, S.~E., \& {Langer}, N. 2010, \apj, 725, 940

\bibitem[{{Yungelson}(1973)}]{Yungelson+1973:1973NInfo..27...93Y}
{Yungelson}, L. 1973, Nauchnye Informatsii, 27, 93

\bibitem[{{Zehe} {et~al.}(2018){Zehe}, {Mugrauer}, {Neuh{\"a}user}, {Pannicke},
  {Lux}, {Bischoff}, {W{\"o}ckel}, \& {Wagner}}]{Zehe+2018:2018AN....339...46Z}
{Zehe}, T., {Mugrauer}, M., {Neuh{\"a}user}, R., {et~al.} 2018, Astronomische
  Nachrichten, 339, 46

\end{thebibliography}

\allowdisplaybreaks
\appendix

\section{Importance of choice of minimum diffusive mixing}\label{app:min_D_mix}

In our \texttt{MESA} models we set a minimum diffusive mixing coefficient, $D_{\rm min}$ (or \texttt{min\_D\_mix} in \texttt{MESA}), in order to account for mixing processes not included in our model and to mix over unphysically sharp composition gradients resulting from the 1D approximation of stellar structure. For our models analysed in this paper we set $D_{\rm min} = 20 \,{\rm cm^2 \, s^{-1}}$, in this Appendix we explore the impact that this choice has on our results.

\begin{figure}[b]
    \centering
    \includegraphics[width=\columnwidth]{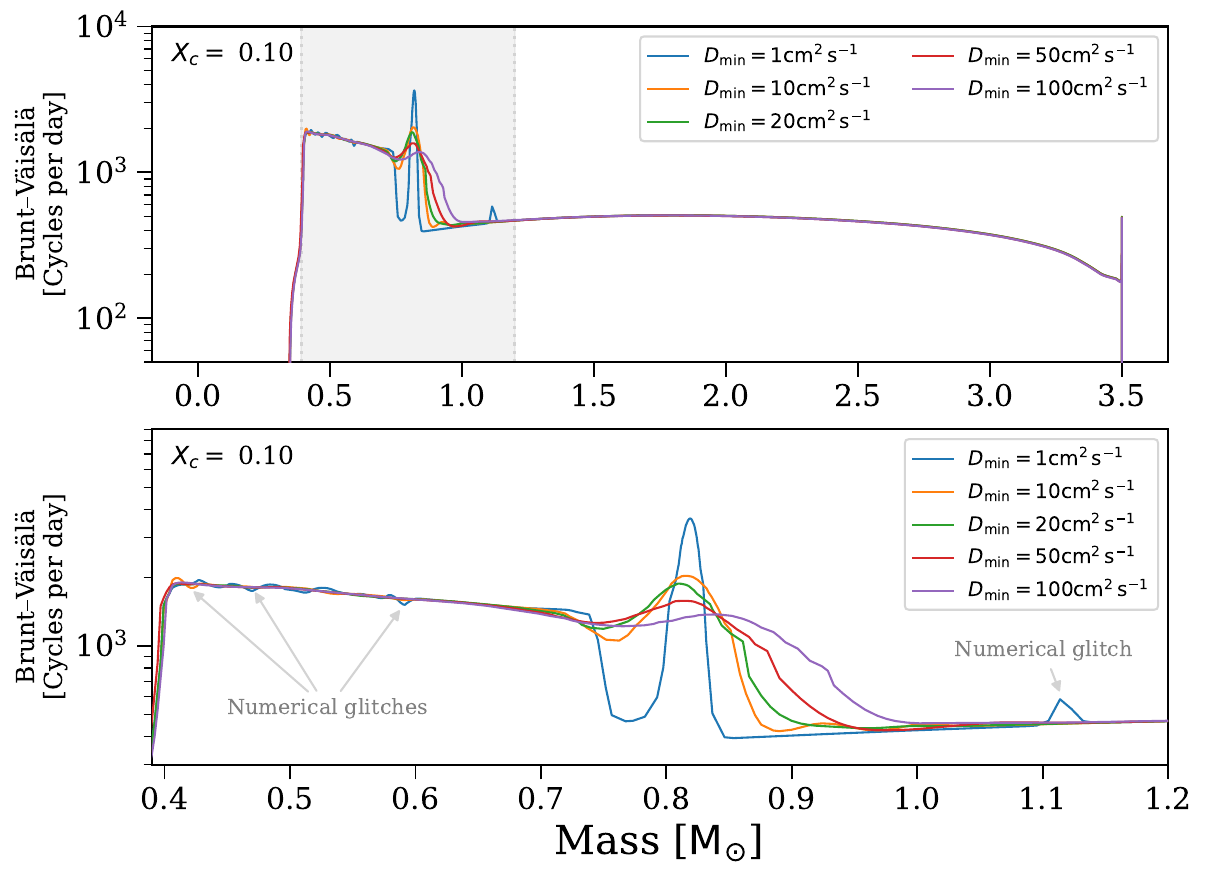}
    \caption{Comparison of the impact of changing the \texttt{MESA} minimum diffusive mixing parameter, $D_{\rm min}$, on the Brunt–Väisälä frequency profile. Bottom panel zooms in on the highlighted range in the top panel. Annotations highlight numerical glitches in low $D_{\rm min}$ models.}
    \label{fig:min_D_mix}
\end{figure}

We repeated our binary \texttt{MESA} simulations for four additional choices of $D_{\rm min}$, ranging from $1 - 100 \,{\rm cm^2 \, s^{-1}}$. In Figure~\ref{fig:min_D_mix} we compare the Brunt–Väisälä frequency profiles of these models at a central hydrogen abundance of $X_c = 0.1$, where the lower panel zooms in on the highlighted region in the upper panel. There are significant differences in the profiles between the different models. As one may expect, lower mixing coefficients lead to steeper composition gradients and therefore sharper features in the Brunt–Väisälä frequency and stronger signals in the period spacing pattern.

However, an overly low choice of $D_{\rm min}$ leads to numerical glitches in the composition gradient and the Brunt–Väisälä frequency profile. These glitches are a result of \texttt{MESA} discretising a 3D gradient in a 1D spherical model, which can lead mesh-point wide spikes in the composition gradient and Brunt–Väisälä frequency, even in higher resolution models. We highlight this in Figure~\ref{fig:min_D_mix}, where glitches are clearly visible in both the $D_{\rm min} = 1 \,{\rm cm^2 \, s^{-1}}$ and $D_{\rm min} = 10 \,{\rm cm^2 \, s^{-1}}$ models.

\begin{figure}[bt]
    \centering
    \includegraphics[width=\columnwidth]{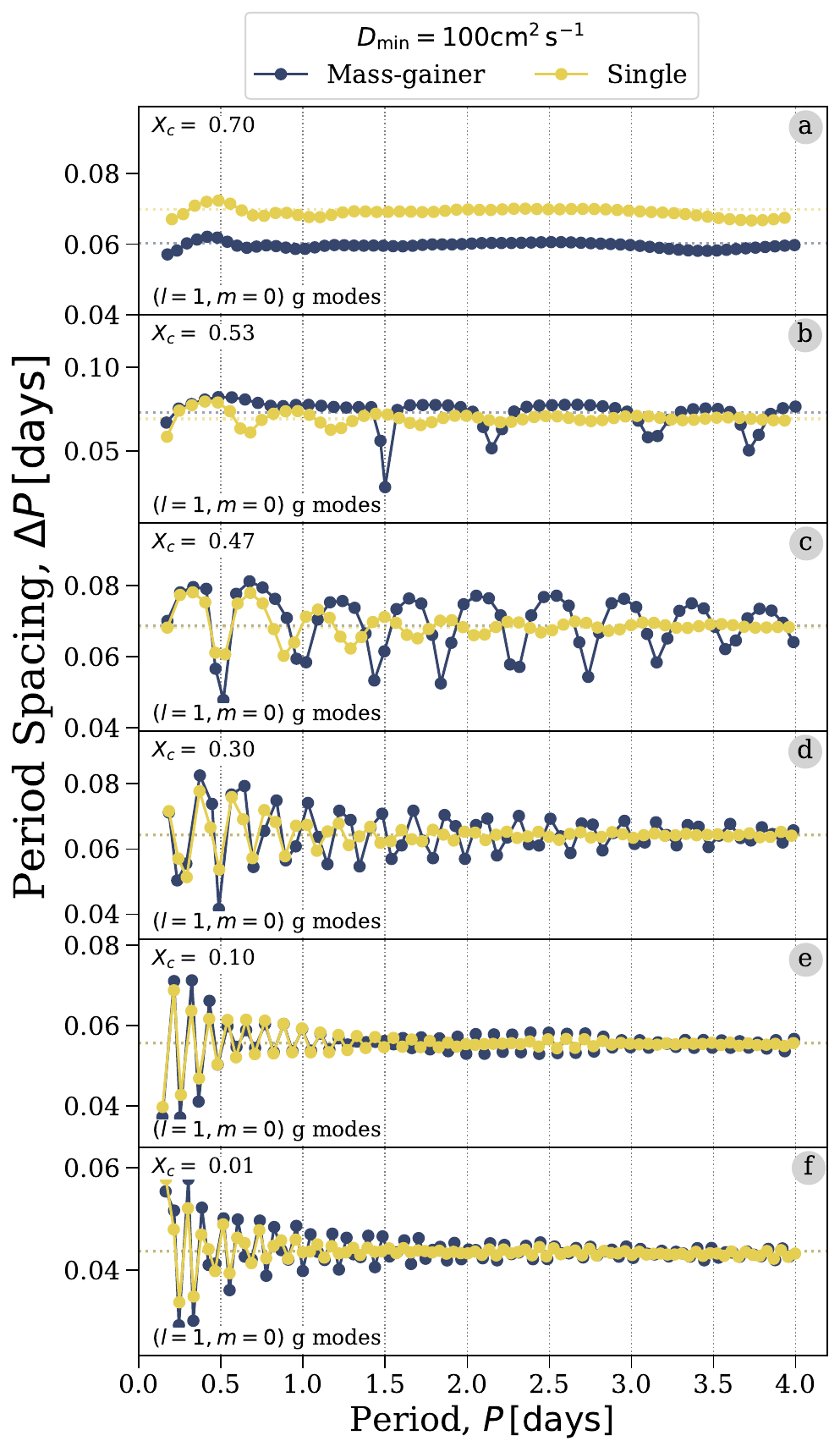}
    \caption{As Figure~\ref{fig:period_spacing}, but with the minimum diffusive mixing coefficient set to $D_{\rm min} = 20 \,{\rm cm^2 \, s^{-1}}$.}
    \label{fig:period_spacing_mdm100}
\end{figure}

We explored a more dense grid of $D_{\rm min}$ models and found that $D_{\rm min} = 20 \,{\rm cm^2 \, s^{-1}}$ was the smallest level of mixing that still removed numerical glitches, which informed our selection of this model as the default choice in this paper.

Many processes are expected to induce extra mixing, such as wave mixing and in particular induced rotation in the accretors \citep{Packet+1981}. While we do not attempt to directly and explicitly model these phenomena, the choice of applying a $D_{\rm min}$ to prevent numerical artifacts partially compensates for this.

We stress that the qualitative differences between the mass-gainer and single star in the period spacing patterns remain the same for all choices of $D_{\rm min}$ that we explored. To highlight this point we show the period spacing pattern for the model with $D_{\rm min} = 100 \,{\rm cm^2 \, s^{-1}}$ in Figure~\ref{fig:period_spacing_mdm100}. Despite slight differences to the exact shape of the pattern, we still find the same features of (i) stronger $\Delta P$ for mass-gainers and (ii) regions in which the period spacing is in-phase and regions in which it is out of phase between the mass-gainer and single star. This confirms that our arbitrary choice of $D_{\rm min}$ does not affect the main conclusions of this study.

\newpage

\section{\texttt{MESA} input physics}\label{app:mesa_inputs}

The MESA EOS is a blend of the OPAL \citep{Rogers2002}, SCVH
\citep{Saumon1995}, FreeEOS \citep{Irwin2004}, HELM \citep{Timmes2000},
PC \citep{Potekhin2010}, and Skye \citep{Jermyn2021} EOSes. Radiative opacities are primarily from OPAL \citep{Iglesias1993, Iglesias1996}, with low-temperature data from \citet{Ferguson2005} and the high-temperature, Compton-scattering dominated regime by \citet{Poutanen2017}. Electron conduction opacities are from \citet{Cassisi2007} and \citet{Blouin2020}. Nuclear reaction rates are from JINA REACLIB \citep{Cyburt2010}, NACRE \citep{Angulo1999} and additional tabulated weak reaction rates \citet{Fuller1985, Oda1994, Langanke2000}.  Screening is included via the prescription of \citet{Chugunov2007}. Thermal neutrino loss rates are from \citet{Itoh1996}. Roche lobe radii in binary systems are computed using the fit of
\citet{Eggleton1983}. For accretors we include thermohaline mixing once they finish accretion following \citet{Kippenhahn+1980} with an efficiency of $\alpha_{\rm thm} = 1$. We follow the \citet{Kolb+1990} prescription for the mass transfer rate in Roche lobe overflowing binary systems with an implicit scheme.

\section{\texttt{MESA} \& \texttt{GYRE} Convergence Tests}\label{app:convergence_tests}

We assessed the numerical convergence of our \texttt{MESA} models by increasing both the number of timesteps and the number of mesh points. Our default model uses \texttt{delta\_mesh\_coeff} $ = 0.4$ and \texttt{delta\_time\_coeff} $ = 1.0$. We decreased both of these by a factor of two (0.2 and 0.5 respectively) and re-ran the analysis for our mass-gainer model.

\begin{figure}
    \centering
    \includegraphics[width=\columnwidth]{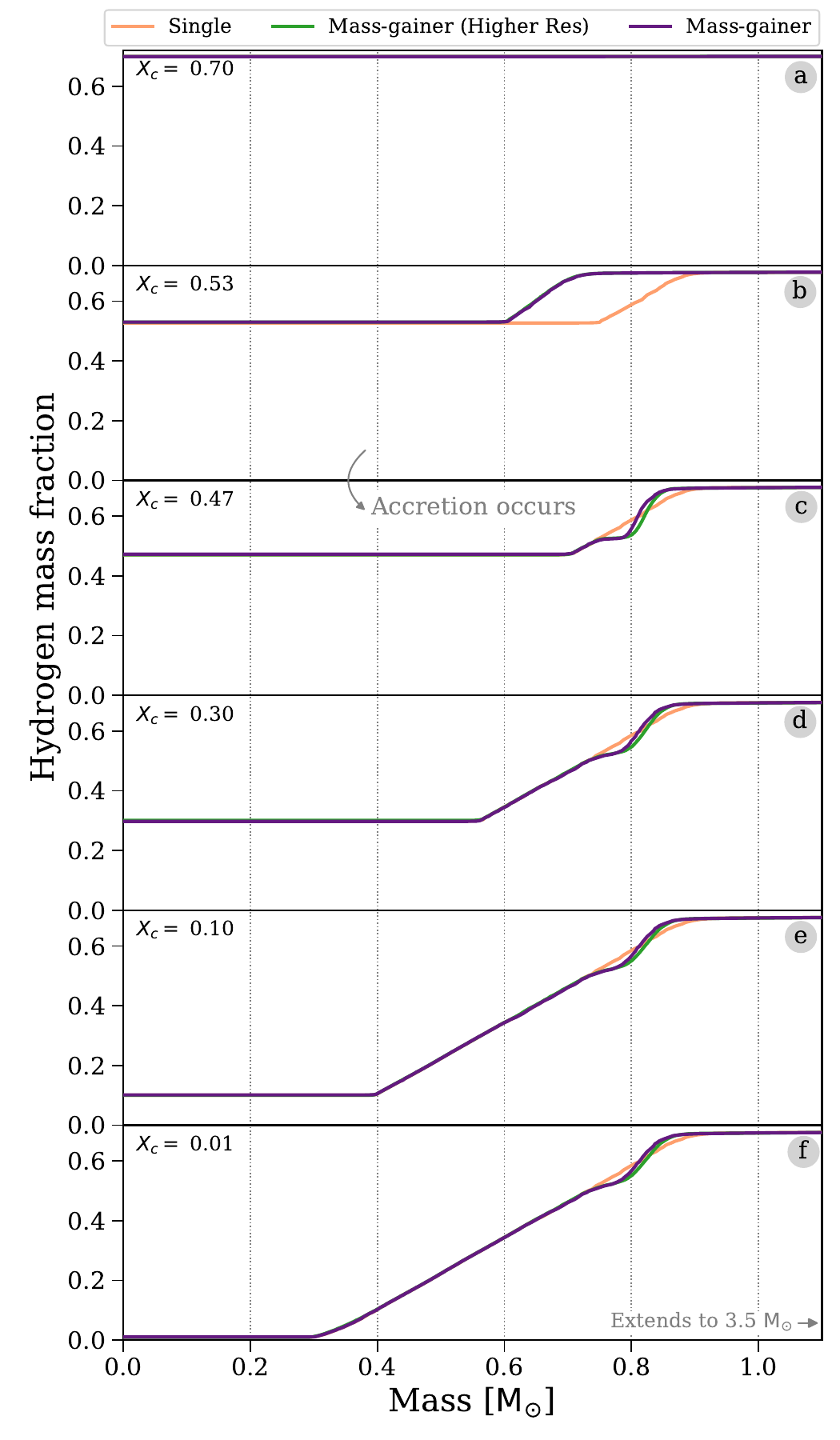}
    \caption{As Figure~\ref{fig:XH_profiles}, but with an additional profile at higher resolution (\texttt{delta\_mesh\_coeff} $ = 0.2$ and \texttt{delta\_time\_coeff} $ = 0.5$) shown in green.}
    \label{fig:xh_convergence}
\end{figure}

\begin{figure}
    \centering
    \includegraphics[width=\columnwidth]{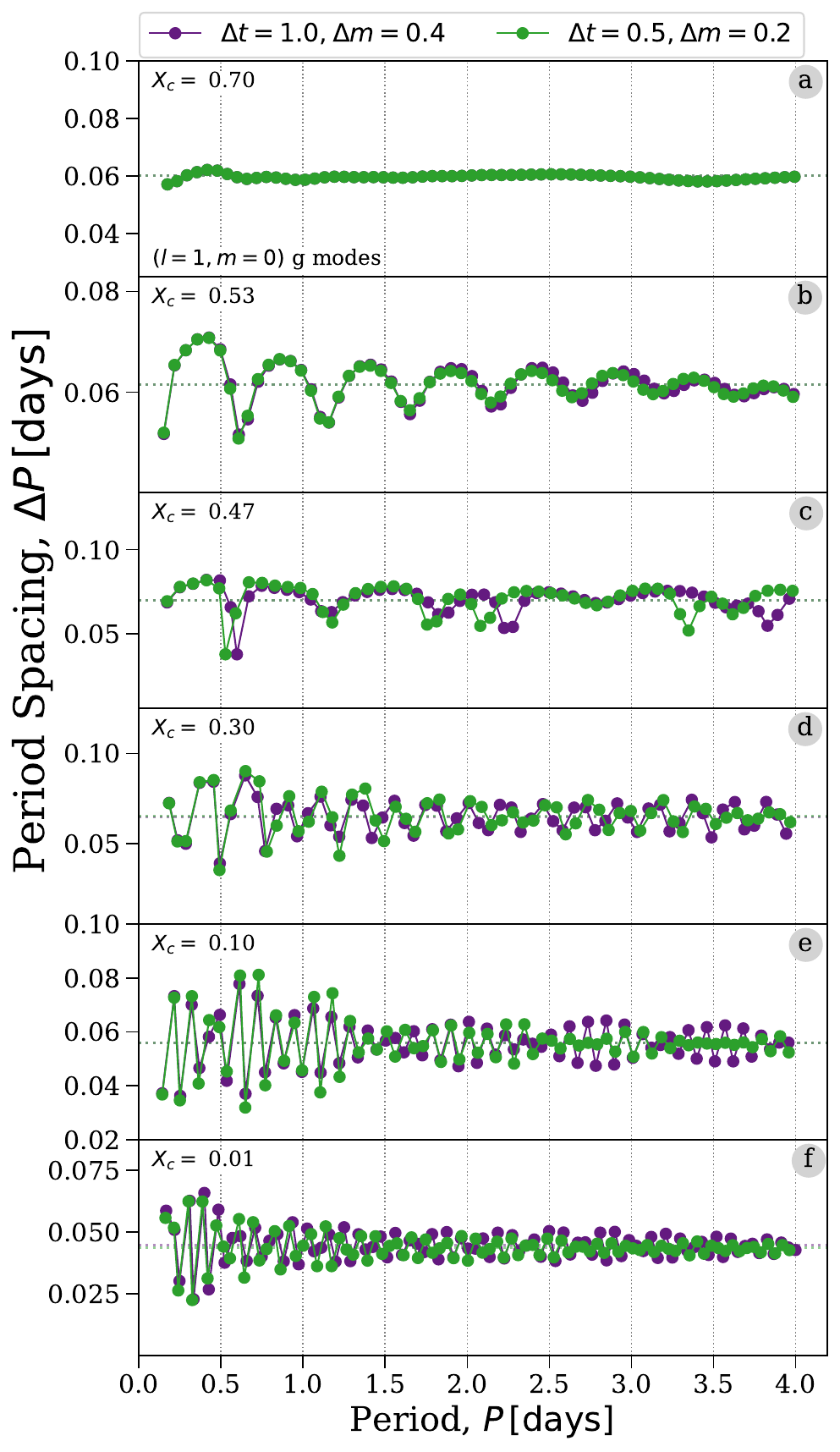}
    \caption{As Figure~\ref{fig:period_spacing}, but now comparing two mass-gainer models at different resolution in timestep and mesh size.}
    \label{fig:period_spacing_convergence}
\end{figure}

In Figure~\ref{fig:xh_convergence} we show the hydrogen abundance profiles with an additional line for the higher resolution model. There are slight changes from our default model, mainly a small offset in mass coordinate, but critically the kink in the distribution is still present throughout the star's evolution. We additionally compare the period spacing pattern for the default and higher resolution models in Figure~\ref{fig:period_spacing_convergence}. The differences here are mainly in terms of amplitude (except in panel c in which a different set of modes are trapped). However, the main shape and features present are still very similar, and compared to the single star model we still see the same key features, primarily in terms of regions with phase offsets after mass transfer. Therefore, given our results are a proof of principle and not quantitative we confirm that our simulations are sufficiently numerically converged and our findings are robust.

For \texttt{GYRE}, as described in Section~\ref{sec:period_spacing}, our frequency scanning range is between 0.25 and 10 cycles per day (equivalent to 0.1 and 4 days) with 2000 steps. Our choice of 2000 steps was determined by iteratively increasing the number of steps until the predicted frequencies of the pulsation modes output by \texttt{GYRE} converged.

\end{document}